\documentclass[%
 reprint,
superscriptaddress,
 amsmath,amssymb,
 aps,
prx,
floatfix,
]{revtex4-1}

\usepackage{ifpdf}
\usepackage{amsmath} 
\usepackage{amssymb}
\usepackage{amsfonts}
\usepackage{braket}
\usepackage{enumitem}
\usepackage{physics}
\usepackage{float}
\usepackage{color}
\usepackage[colorlinks=true,linkcolor=blue,citecolor=blue,breaklinks]{hyperref}
\usepackage{graphicx}
\usepackage{dcolumn}
\usepackage{bm}
\usepackage[normalem]{ulem}
\usepackage{mathtools}
\usepackage{verbatim}
\usepackage{xcolor}

\newcommand{\up}{{\uparrow}}
\newcommand{\dn}{{\downarrow}}
\newcommand{\hc}{\text{H.c.}}

\usepackage{tikz}
\usetikzlibrary{math}

\definecolor{Zcolour}{rgb}{0.992, 0.588, 0.22}
\definecolor{purple}{rgb}{0.5,0,0.5}
\definecolor{brown}{rgb}{0.6,0.2,0}
\definecolor{dkgreen}{rgb}{0,0.5,0}

\begin{document}


\title{Eta-pairing states as true scars in an extended Hubbard Model}
\author{Daniel K.~Mark}
\affiliation{Department of Physics, California Institute of Technology, Pasadena, California 91125, USA}
\author{Olexei I.~Motrunich}
\affiliation{Department of Physics, California Institute of Technology, Pasadena, California 91125, USA}

\date{\today}

\begin{abstract}
The eta-pairing states are a set of exactly known eigenstates of the Hubbard model on hypercubic lattices, first discovered by Yang [Phys.~Rev.~Lett.~\textbf{63}, 2144 (1989)]. These states are not many-body scar states in the Hubbard model because they occupy unique symmetry sectors defined by the so-called ``eta-pairing SU(2)" symmetry. We study an extended Hubbard model with bond-charge interactions, popularized by Hirsch [Physica C \textbf{158}, 326 (1989)], where the eta-pairing states survive without the eta-pairing symmetry and become true scar states. We also discuss similarities between the eta-pairing states and exact scar towers in the spin-1 XY model found by Schecter and Iadecola [Phys.~Rev.~Lett.~\textbf{123}, 147201 (2019)], and systematically arrive at all nearest-neighbor terms that preserve such scar towers in 1D. We also generalize these terms to arbitrary bipartite lattices. Our study of the spin-1 XY model also leads us to several new scarred models, including a spin-1/2 $J_1-J_2$ model with Dzyaloshinkskii-Moriya interaction, in realistic quantum magnet settings in 1D and 2D.
\end{abstract}

\maketitle

\section{Introduction}
Quantum many-body scar states refer to sets of exceptional states in the spectra of some many-body non-integrable models. These states do not obey the Eigenstate Thermalization Hypothesis (ETH)~\cite{Deutsch1991,Srednicki1994}, a framework used to describe how closed quantum many-body systems equilibriate to thermal distributions for local measurements. Scar states violate the ETH by having local quantities that are different from those of nearby states in energy, and in particular by having sub-volume law entanglement entropy (EE) scaling. Unlike many-body localized systems~\cite{Basko2006,Oganesyan2007:MBLhighT,Bardarson2012:UnboundedEEgrowthMBL,Serbyn2013:SlowEEgrowth,serbyn_local_2013,Huse2014:PhenomMBL,Bauer2013:AreaLawMBL,nandkishore_many-body_2015,Abanin2019:RMPonMBL}, where \emph{every} eigenstate violates the ETH, quantum many-body scarred systems are an interesting class of models where only a small number of eigenstates violate the ETH in an otherwise thermal spectrum. 

Quantum many-body scarring was first observed in a cold Rydberg atom experiment~\cite{Bernien2017}. The experiment is modelled by the ``PXP model"~\cite{turner_quantum_2018,Turner2018}. While there are several approximate ways of understanding the scar states in the PXP model~\cite{Bernien2017, turner_quantum_2018, Turner2018, khemani_signatures_2018, ho_periodic_2019, Lin2019, Choi2019, surace_lattice_2019, Iadecola2019, Shiraishi2019:PXPembeddedH, michailidis_slow_2019, moudgalya_quantum_2019, lin_slow_2019, bull_quantum_2020, michailidis_stabilizing_2020, lin_quantum_2020}, only some eigenstates in the middle of the spectrum are known exactly~\cite{Lin2019, Shiraishi2019:PXPembeddedH}.

Conversely, there are non-integrable systems with exactly known scar states, such as the AKLT model~\cite{Moudgalya2018, moudgalya_entanglement_2018, moudgalya_large_2020}, the spin-1 XY model~\cite{schecter_weak_2019, chattopadhyay_quantum_2019} and a spin-1/2 domain-wall-conserving model~\cite{iadecola_quantum_2019,mark_unified_2020}, among others~\cite{shiraishi_systematic_2017, Mori2017:ThermalizationWithoutETH, James2019:Confinement, khemani_local_2019, hudomal_quantum_2019, pancotti_quantum_2019, Pai2019:ScarsFracton, Sala2019:Fragmentation,shibata_onsagers_2019}, including a framework to make target states as scars in non-integrable models~\cite{shiraishi_systematic_2017, Mori2017:ThermalizationWithoutETH}; there is also a growing number of examples of scars in the Floquet setting~\cite{Pai2019:ScarsFracton, mukherjee_collapse_2019, haldar_scars_2019, sugiura_many-body_2019, zhao_quantum_2020, mizuta_exact_2020}.

The Hubbard model stands apart from this list of models. There are exactly known states embedded in the spectrum---the eta-pairing states due to Yang~\cite{yang__1989}. These eta-pairing states are known to have sub-volume law entanglement~\cite{vafek_entanglement_2017}. However, they do not constitute many-body scar states because the Hubbard model possesses an additional SU(2) symmetry~\cite{Yang_SO4_1990}, which we refer to as the ``eta-pairing SU(2)." The eta-pairing states are the unique eigenstates in the symmetry sector of maximal ``eta-pairing SU(2)" total spin and are therefore not expected to be thermalized with respect to the rest of the spectrum.  While there are several proposals for unusual thermalization in the Hubbard model~\cite{garrison_partial_2017, veness_quantum_2017, veness_atypical_2017, yu_beyond_2018, iadecola_exact_2019}, in this paper we discuss a direct way to make an electronic Hubbard-like model quantum scarred by the eta-pairing states.

We will present several terms that break the eta-pairing SU(2) symmetry, but preserve the eta-pairing states $\ket{\psi_N}$ (defined below) as eigenstates. This can be achieved while retaining the spin SU(2) symmetry,
so $\ket{\psi_N}$ will be eigenstates in the otherwise thermal sector with quantum numbers total spin $S = 0$, momentum $k = N \pi~ (\text{mod } 2\pi)$, site inversion $I_s = 1$, fermion species numbers $N_\uparrow = N_\downarrow = N$,
and thus constitute many-body scar states. The most notable of these perturbations is the Hirsch model discussed below. 

In Section~\ref{sec:Hirsch} we review the Hubbard model and the Hirsch model, and discuss how the eta-pairing states (and some related states) are scars in this model. In order to discuss our systematic construction of terms that make the eta-pairing states scars, in Section~\ref{sec:relationbetweenfermionandspin} we first draw parallels between spin-1 and electronic models, in particular between the spin-1 XY model scars and the eta-pairing states. In Section~\ref{sec:spinmodels} we systematically construct such scarred models in spin-1 systems, which goes beyond previously known models. We then map these results over to electronic systems in Section~\ref{sec:translationtofermion}, with the result that the Hirsch model belongs to a small family of models scarred by the eta-pairing states, and is arguably the most easily realized of this family.

Lastly, our study of the spin-1 XY model scars with $k=\pi$ bimagnon towers naturally leads us to consider $k=0$ bimagnon towers as scars. In Appendix~\ref{sec:q0towers}, we show that the spin-1 $k=0$ bimagnon towers are scars in a model with the Dzyaloshinskii-Moriya interaction (DMI) replacing the XY spin-exchange term. In the presence of conservation of the number of 0s (i.e., sites where $S_j^z=0$), the $k=0$ bimagnon tower maps onto a spin-1/2 $k=0$ magnon tower. We also find a model scarred by this tower: the $J_1-J_2$ model with spin-1/2 DMI.

\section{The Hirsch Model: Hubbard Model with Bond-Charge Interactions}
\label{sec:Hirsch}
\subsection{The Hubbard Model, eta-pairing states and eta-pairing SU(2) symmetry}
The Hubbard model is a model of interacting electrons given by
\begin{multline}
    H_\text{Hub.} = -t \sum_{\langle \mathbf{ij} \rangle, \sigma} \left(c^\dagger_{\mathbf{i},\sigma} c_{\mathbf{j},\sigma}^{\mathstrut} + \hc \right) \\
    + U \sum_\mathbf{j} n_{\mathbf{j}, \uparrow} n_{\mathbf{j}, \downarrow} - \mu \sum_{\mathbf{j}, \sigma} n_{\mathbf{j}, \sigma} ~,
\end{multline}
where $c_{\mathbf{j},\sigma}^\dagger$ creates an electron at site $\mathbf{j}$ with spin $\sigma$, and $n_{\mathbf{j}, \sigma} = c_{\mathbf{j},\sigma}^\dagger c_{\mathbf{j},\sigma}$.
The model comprises of hopping, interaction and chemical potential terms with coefficients $t, U$ and $\mu$ respectively.  For analysis in later sections, it will be convenient to denote the on-site states as $\{h, \uparrow, \downarrow, d\}$, where $h$ indicates an empty site (``holon"), $\uparrow$ or $\downarrow$ indicates a singly occupied site with a spin-up or spin-down electron, and $d$ indicates a doubly occupied site (``doublon").
The Hubbard model on the 1D chain with nearest-neighbor hopping is exactly solved~\cite{essler_frahm_gohmann_klumper_korepin_2005}, but this is not important for the present work. 

Yang~\cite{yang__1989} introduced the eta-pairing states as eigenstates of the Hubbard model on any bipartite lattice. For our purposes we will specialize to hypercubic lattices of $V$ sites. The eta-pairing states are:
\begin{equation}
    \ket{\psi_N} = C_N \left(\eta^\dagger \right)^N \ket{\text{vac.}}, \label{eq:eta pairing states}
\end{equation}
where
\begin{equation}
    \eta^\dagger = \sum_{\mathbf{j}} e^{i \mathbf{r_j} \cdot \pmb{\pi}} c^\dagger_{\mathbf{j}, \uparrow} c^\dagger_{\mathbf{j}, \downarrow} ~,
    \label{eq:etadagger}
\end{equation}
where $C_N = \sqrt{(V-N)!/(V!N!)}$ is the normalization constant, $\ket{\text{vac.}}$ is the electronic vacuum state and $\pmb{\pi} = (\pi, \pi, \dots, \pi)$.  The number of pairs $N$ can range from 0 to $V$. We also note that our definition of $\eta^\dagger$ differs from that in the literature by a sign --- this choice is made for easy comparison with the spin-1 model operator $Q^\dagger$ below. There is a significant body of work investigating the possibility of realizing these states as ground states as well as signatures of eta-pairing in Hubbard models, see e.g. Refs.~\cite{shen_exact_1993,de_boer_exact_1995,kitamura_eta-pairing_2016,li_exact_2019,li_long-range_2019,kaneko_photoinduced_2019}.

$\ket{\psi_N}$ has energy $(U - 2\mu)N$. It is trivially an eigenstate of the interaction and chemical potential terms, and is annihilated by the hopping term, owing to cancellations from the momentum $\pmb{\pi}$ construction~\cite{yang__1989}.

When $U = 2\mu$, $H_\text{Hub.}$ in fact is SU(2) symmetric under the following generators~\cite{Yang_SO4_1990}: $[H_\text{Hub.}, \eta^\dagger] = [H_\text{Hub.}, \eta] = [H_\text{Hub.}, \eta^z] = 0$, where 
\begin{equation}
    \eta^z = \frac{1}{2} \left(\sum_\mathbf{j,\sigma} n_{\mathbf{j},\sigma} - V \right) = \frac{1}{2}(N_\up + N_\dn
    - V) ~.
\label{eq:etaz}
\end{equation}
$H_\text{Hub.}$ is also SU(2) symmetric under the total electron spin generators $S^\pm, S^z$, where 
 \begin{align}
     & S^+ = \sum_\mathbf{j} c^\dagger_{\mathbf{j},\uparrow} c_{\mathbf{j},\downarrow}^{\mathstrut} ~, \quad S^- = (S^+)^\dagger ~, \\
     & S^z = \frac{1}{2} \sum_\mathbf{j} \left(n_{\mathbf{j},\uparrow} -  n_{\mathbf{j},\downarrow} \right) = \frac{1}{2}(N_\uparrow - N_\downarrow) ~.
 \end{align}
The generators $\eta^\dagger, \eta, \eta^z$ and $S^\pm, S^z$ obey su(2) commutation relations $[\eta^z, \eta^\dagger] = \eta^\dagger,~[\eta^\dagger, \eta] = 2\eta^z$, and all eta-pairing generators commute with all spin SU(2) generators. Therefore there are two independent SU(2) symmetries in $H_\text{Hub.}$. Changing $\mu$ simply shifts the energies by the corresponding values determined by the $\eta^z$ quantum numbers and does not affect the eigenstates or change any symmetry sectors.

The eta-pairing states $\ket{\psi_N}$ lie in the spin sector $S=0$. They have $\eta^z = (2N - V)/2$.
Therefore, as $N$ runs from $0$ to $V$, they comprise the unique multiplet of states with the maximum possible total eta-pairing ``spin" $V/2$.

On any lattice, at fixed pair
density $\nu = N/V$, these states were shown in Ref.~\cite{vafek_entanglement_2017} to have sub-volume law entanglement entropy:
\begin{equation}
    S_A = \frac{1}{2} (1 + \ln[2\pi \nu (1-\nu) V_A] ) ~, 
\end{equation}
where $S_A$ is measured at a cut which partitions $V$ into $V_A$ and $V-V_A$ sites. However, because $\ket{\psi_N}$ are the unique states in a given $\eta^z$, $\eta^2$ sector, Ref.~\cite{vafek_entanglement_2017} concluded that they did not violate the ETH.

Lastly, we note that other related exact states are known in the Hubbard model~\cite{Yang_SO4_1990, vafek_entanglement_2017}:
\begin{equation}
    \ket*{\psi_N^{\{\mathbf{k}\})}} = C_N^{\{\mathbf{k}\}} (\eta^\dagger)^N \prod_{\mathbf{k} \in \mathcal{F}} c^\dagger_\downarrow(\mathbf{k}) \ket{\text{vac.}},
\end{equation}
for any set $\mathcal{F}$ of wavevectors $\mathbf{k}$, because there is no interaction between electrons of the same spin species. However, we restrict our attention to the original eta-pairing states $\ket{\psi_N}$ because they are the only states that survive under the Hirsch term and other pertubations (and in any dimension).

\subsection{The Hirsch Model}

\begin{figure*}
\centering
\includegraphics[width=\textwidth]{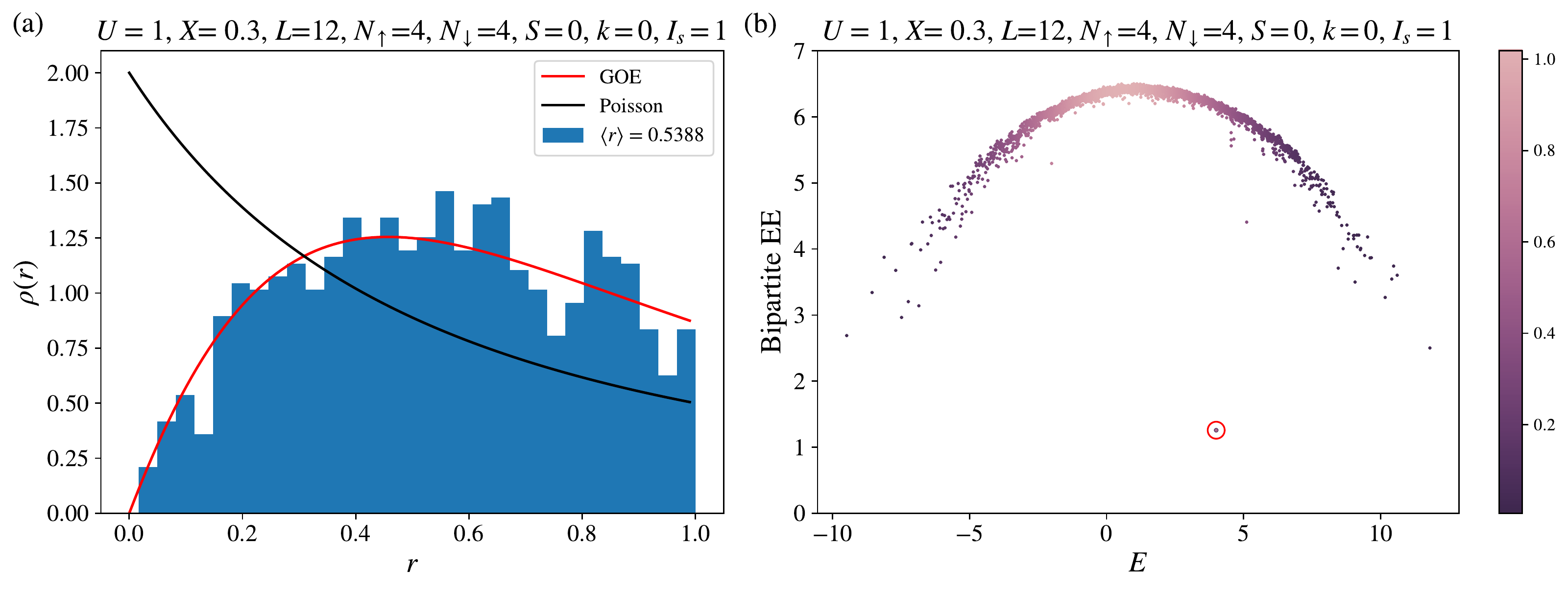}
\caption{(a) r-statistics of the Hirsch model with $U=t=1$ and $X=0.3$ on a periodic chain of length $L = 12$, in the symmetry sector $N_\uparrow = N_\downarrow = 4$, total spin $S=0$, momentum $k=0$ and site inversion $I_s = 1$. The Hilbert space dimension is 3072. The r-statistics are consistent with the Wigner-Dyson GOE prediction for quantum chaotic models. (b) Bipartite entanglement entropy (EE) in the same symmetry sector. The eta-pairing state $\ket{\psi_4}$ is a clear EE outlier. Each point is colored by the density of states at the corresponding energy, with colorbar (normalized to 1) on the right.
}
\label{fig:HubbardED}
\end{figure*}

The Hirsch model was popularized by Hirsch~\cite{hirsch_bond-charge_1989, hirsch_superconducting_1989} as a model of hole superconductivity. It is given by
\begin{multline}
    H =  -\sum_{\langle \mathbf{ij} \rangle, \sigma} \left[t - X (n_{\mathbf{i},-\sigma} + n_{\mathbf{j},-\sigma}) \right]  \left(c^\dagger_{\mathbf{i},\sigma} c_{\mathbf{j},\sigma}^{\mathstrut} + \hc \right)  \\
+ U \sum_\mathbf{j} n_{\mathbf{j}, \uparrow} n_{\mathbf{j},\downarrow} - \mu \sum_{\mathbf{j},\sigma} n_{\mathbf{j},\sigma} = H_\text{Hub.} + H_\text{Hirsch} ~,
\label{eq:Hirschmodel}
\end{multline}
where ``$-\sigma$" indicates the opposite spin species to $\sigma$. The Hirsch model adds to the Hubbard model a ``bond-charge interaction" term~\cite{Micnas_superconductivity_1989, Micnas_superconductivity_RMP_1990,strack_hubbard_1993, de_boer_ensurematheta_1995, de_boer_exact_1995}, which modifies the hopping constant depending on the occupations across the bond. This term is alternatively called ``correlated hopping"~\cite{bulka_superconductivity_1998, vidal_strongly_2001}. This term was originally estimated by Hubbard~\cite{hubbard_electron_1963} in solid state systems, and there are also proposals to realize this with ultracold atoms~\cite{Duan_2007}. We lastly note that the 1D Hirsch model is integrable when $X = t$~\cite{arrachea_exact_1994,Schadschneider_superconductivity_1995}, but this special point is not important for the scar physics of interest to us.

The Hirsch term breaks the eta-pairing SU(2) symmetry, which can be verified by evaluating the commutator $[H_\text{Hirsch}, \eta^\dagger]$. However, the Hirsch term preserves the spin SU(2). This is clear by the re-writing on each bond $\langle\mathbf{ij}\rangle$:
\begin{multline}
    \sum_\sigma (n_{\mathbf{i},-\sigma} + n_{\mathbf{j},-\sigma}) \left(c^\dagger_{\mathbf{i},\sigma} c_{\mathbf{j},\sigma}^{\mathstrut} + \hc \right)  \\
    = \left[\sum_{\sigma'} (n_{\mathbf{i},\sigma'} + n_{\mathbf{j},\sigma'}) - 1 \right] \sum_\sigma \left(c^\dagger_{\mathbf{i},\sigma} c_{\mathbf{j},\sigma}^{\mathstrut} + \hc \right) ~. \label{eq:Hirschrewrite} 
\end{multline}
From Eq.~(\ref{eq:Hirschrewrite}), we also immediately see that $H_\text{Hirsch} \ket{\psi_N} = 0$, because on bonds connecting opposite sublattices, the right-most operator in Eq.~(\ref{eq:Hirschrewrite}) annihilates $\ket{\psi_N}$: $\sum_\sigma \left(c^\dagger_{\mathbf{i},\sigma} c_{\mathbf{j},\sigma}^{\mathstrut} + \hc \right) \ket{\psi_N} = 0$. The latter is the property from Yang's original paper~\cite{yang__1989} and is easily verified because the only non-trivial term is:
\begin{align}
    &\sum_\sigma \left(c^\dagger_{\mathbf{i},\sigma} c_{\mathbf{j},\sigma}^{\mathstrut} + \hc \right) \left(c^\dagger_{\mathbf{i},\up}c^\dagger_{\mathbf{i},\dn} - c^\dagger_{\mathbf{j},\up}c^\dagger_{\mathbf{j},\dn} \right) \ket{\text{vac.}} \\
    &= \left(c^\dagger_{\mathbf{j},\up}c^\dagger_{\mathbf{i},\dn} - c^\dagger_{\mathbf{j},\dn}c^\dagger_{\mathbf{i},\up} - c^\dagger_{\mathbf{i},\up}c^\dagger_{\mathbf{j},\dn} + c^\dagger_{\mathbf{i},\dn}c^\dagger_{\mathbf{j},\up} \right) \ket{\text{vac.}} = 0 ~. \nonumber 
\end{align}
Therefore the eta-pairing states $\ket{\psi_N}$ remain eigenstates of $H$ for any strength of the Hirsch term.
They are embedded in the symmetry sectors $S=0,~\mathbf{k} = N \pmb{\pi}~ (\text{mod } 2\pmb{\pi}),~I_s = 1,~N_\uparrow = N_\downarrow = N$, where $S$ denotes total spin, $\mathbf{k}$ momentum, $I_s$ site inversion, and $N_\sigma$ the number of $\sigma$ species electrons.

From numerical studies of the above model in 1D systems with periodic boundary conditions (PBCs), we verify that the level-spacing statistics in these sectors follow the Wigner-Dyson GOE predictions, see Fig.~\ref{fig:HubbardED}(a), and that $\ket{\psi_N}$ are bipartite EE outliers in their respective sectors, see Fig.~\ref{fig:HubbardED}(b).

We therefore conclude that in the Hirsch model, the eta-pairing states constitute true many-body scar states. We finally note that while the Hirsch model has been around for some time, the presence of the scar states in the non-integrable model was not known.  Also, we identified it through a systematic study of two-site terms that break the eta-pairing SU(2) but preserve the eta-pairing states, discussed in Section~\ref{sec:translationtofermion} below.

Lastly, while we identified the Hirsch model as a model that contains the eta-pairing states as scars, our subsequent numerical investigation revealed additional entanglement outlier states in 1D, some of which constitute two additional scar towers $\ket{\psi_{N,M}} = (t^\dagger)^M \ket{\psi_N}$ and $\ket{\phi_{N}} = s^\dagger \ket{\psi_N}$. The operators $t^\dagger$ and $s^\dagger$ create nearest-neighbor triplets $\ket{\uparrow,\uparrow}$ and singlets $\ket{\uparrow,\downarrow} - \ket{\downarrow,\uparrow}$, with momentum $\pi$. They are defined in Appendix~\ref{sec:AppendixTriplSinglScars}, in which we also prove these states.

\section{Analogy between Spin-1 XY scar tower and Eta-pairing states}
\label{sec:relationbetweenfermionandspin}
To systematically construct Hamiltonians that make the eta-pairing states scarry, we first understand scarring in spin-1 systems. In particular, the spin-1 XY model~\cite{schecter_weak_2019, chattopadhyay_quantum_2019} is known to contain an exact tower of scar states, which are analogous to the eta-pairing states in the Hubbard model. We first construct scarred models in the simpler spin-1 setting, then show how these results translate to electronic models. We also note that while there is a separate tower of scar states in the \mbox{spin-1} AKLT model~\cite{Moudgalya2018,moudgalya_entanglement_2018}, their electronic model analogues are not immediate.

The spin-1 XY model is one of the simplest models known to have an exact tower of scar states~\cite{schecter_weak_2019, chattopadhyay_quantum_2019}. In Ref.~\cite{schecter_weak_2019}, Schecter and Iadecola considered spin-1 XY models of the form:
\begin{align}
    H_\text{Spin-1} &= J\sum_{\expval{\mathbf{ij}}} (S_\mathbf{i}^x S_\mathbf{j}^x + S_\mathbf{i}^y S_\mathbf{j}^y) + h \sum_\mathbf{j} S_\mathbf{j}^z + D \sum_\mathbf{j} \left(S_\mathbf{j}^z \right)^2 \nonumber \\
    &= H_{XY} + H_{z} + H_{z^2} ~.
    \label{eq:HIS}
\end{align}
Here the $\mathbf{S}_\mathbf{i}$ are spin-1 spin operators. In what follows whether the $S$'s refer to spin operators of the spin-1 model or electron spin operators of the Hubbard model should be clear from context.

In 1D, one has to introduce a third-neighbour term $H_3 = J_3 \sum_j (S_j^x S_{j+3}^x + S_j^y S_{j+3}^y)$ to break a special non-local SU(2) symmetry present in sectors with even magnetization~\cite{kitazawa_ansu2_2003, chattopadhyay_quantum_2019}. On hypercubic lattices, the scar tower $\ket{\mathcal{S}_N}$ is formed by the operator $Q^\dagger$,
\begin{equation}
    \ket{\mathcal{S}_N} = \left(Q^\dagger \right)^N \ket{\Omega}, \quad Q^\dagger = \frac{1}{2}\sum_\mathbf{j} e^{i \mathbf{r_j} \cdot \pmb{\pi}} \big(S_\mathbf{j}^+ \big)^2 ~,
    \label{eq:ISXYscar1}
\end{equation}
where $\ket{\Omega} = \ket{-1,-1,...,-1}$ and $N = 0,...,V$. These states have energies $E_N = h(2N-V) + DV$.

We immediately see that the $Q^\dagger$ operator in the spin-1 model is analogous to the $\eta^\dagger$ operator in the Hubbard model [Eq.~(\ref{eq:etadagger})]. $\frac{1}{2}(S^+)^2$ sends $\ket{-1} \rightarrow \ket{1}$, while $c^\dagger_\uparrow c^\dagger_\downarrow$ sends $\ket{h} \rightarrow \ket{d}$, where $h$ indicates an empty site (``holon") and $d$ indicates a doubly occupied site (``doublon"). This hints at identifying the spin-1 states $\ket{-1},\ket{1}$ with the electronic $\ket{h},\ket{d}$ respectively. Furthermore, comparison of su(2) algebra relations such as $[Q^\dagger, Q] = S^z_\text{Tot.}$ in the spin-1 case with $[\eta^\dagger, \eta] = N_\up + N_\dn - V$ in the electron case suggests relating spin-1 models with $S^z_\text{Tot.}$ conservation and electronic models with conserved total electron number. There is some ambiguity with identification of the spin-1 state $\ket{0}$, but we will argue that for the purposes of our study, electron spin SU(2)-invariance---which is natural to require in an electronic model to reduce the space of models under consideration---makes our identification unambiguous.

\section{Systematic construction of spin-1 models with the XY scar tower}
\label{sec:spinmodels}
In Ref.~\cite{schecter_weak_2019}, Schecter and Iadecola noted an SU(2) symmetry obeyed only by the scar tower $\ket{\mathcal{S}_N}$. In Ref.~\cite{mark_unified_2020}, we found a model ``embedded" in the XY model:
\begin{equation}
    H_0 = \sum_{\expval{\mathbf{ij}}} \left(\ketbra{1,0}{0,1} - \ketbra{-1,0}{0,-1} + \hc \right)_{\mathbf{i,j}} ~, \label{eq:H0}
\end{equation}
such that
\begin{equation}
    H_{XY} = H_0 + H'_{XY} ~, \label{eq:XYdecomp}
\end{equation}
where
\begin{multline}
    H'_{XY} = \sum_{\expval{\mathbf{ij}}}
    \Big[\big(\ket{1,-1} + \ket{-1,1}\big)\! \bra{0,0} \\
    + 2\ketbra{-1,0}{0,-1} + \hc \Big]_{\mathbf{i,j}} ~. 
\end{multline}
This rewriting is significant because $H_0$ commutes with $Q^\dagger$ and $S^z_\text{Tot.}$, while $H'_{XY}$ annihilates $\ket{\mathcal{S}_N}$. Operators $Q^\dagger, Q$, and $Q^z \equiv S^z_\text{Tot.}/2$ obey the su(2) commutation relations and therefore generate a ``pseudospin" SU(2) symmetry present in $H_0$ analogous to the eta-pairing SU(2) in the Hubbard model. As with the eta-pairing states, $\ket{\mathcal{S}_N}$ are the unique states occupying the sector of the highest total pseudospin $V/2$. The pseudospin SU(2) is broken by $H'_{XY}$, and hence the total Hamiltonian $H_{XY}$ contains $\ket{\mathcal{S}_N}$ in an otherwise thermal spectrum.

The above observation inspired this work, and a natural question is to systematically generate all other spin-1 nearest-neighbor models that share the scar tower $\ket{\mathcal{S}_N}$. To do so we list the following operators:
\begin{enumerate}
    \item[A.] Nearest-neighbor operators that commute with $Q^\dagger$ and $S^z_\text{Tot.}$.
    \item[B.] Nearest-neighbor operators $O$ such that $[O, Q^\dagger] = Q^\dagger$.
    \item[C.] Nearest-neighbor operators that annihilate $\ket{\mathcal{S}_N}$.
\end{enumerate}
For simplicity, in each group we only consider Hermitian operators, although our procedure is also able to find non-Hermitian operators satisfying the above properties.

It is clear that if $\ket{\Omega}$ is an eigenstate of a model in group A, so too are all the $\ket{\mathcal{S}_N}$.  This is immediately true for the state $\ket{\Omega}$, because this state has the highest possible pseudospin quantum number of $V/2$ and hence is always an eigenstate of any model with the corresponding SU(2) symmetry. 

It is natural to consider terms in groups B and C modulo those in group A. As discussed below, it turns out that the unique operator in group B, modulo terms in group A, is $Q^z = S^z_\text{Tot.}/2$. Adding a term proportional to $Q^z$ to the Hamiltonian uniformly shifts each $S^z_\text{Tot.}$ sector in  energy, but does not change any eigenvectors or symmetry sectors. In any model comprised of terms from groups A and B, the Iadecola-Schecter states $\ket{\mathcal{S}_N}$ will not be ``true" scar states for the same reasons that the eta-pairing states are not in the original Hubbard model.

We can break the pseudospin symmetry and turn $\ket{\mathcal{S}_N}$ into true scars with any linear combination of such a model and those in group C that annihilate $\ket{\mathcal{S}_N}$. In the example in Eq.~(\ref{eq:XYdecomp}), $H_0$ belongs to group A, and $H'_{XY}$ to group C. The energies of the scar states are split by the term $H_z = h S^z_\text{Tot.}$.

\subsection{Operators that commute with $Q^\dagger$ and $S^z_\text{Tot.}$}
\label{subsec:opsthatcommute}
We first note that $[O, S^z_\text{Tot.}] = 0$ is a necessary condition for $[O, Q^\dagger] = 0$, because $S^z_\text{Tot.} = [Q^\dagger, Q]$.

It is immediate that the only on-site operators that commute with $Q^\dagger$ are $\ketbra{0}_j$ and $\ketbra{1}_j + \ketbra{-1}_j = (S^z_j)^2$ (the two terms sum to identity, so are not independent for our purposes). We list these operators in Table~\ref{tab:table1} for reference. We also list their translations to electronic models, which will be explained later in Sec.~\ref{sec:translationtofermion}.

\begin{table}
    \centering
    \begin{tabular}{c c} \toprule
    \textbf{Spin-1 model} & \textbf{Electronic model} \\
    \colrule
    \multicolumn{2}{c}{\textbf{A. Operators that commute with $Q^\dagger$, or with $\eta^\dagger$} }\\
    \colrule
    $\ketbra{0}_j$ & $\ketbra{\uparrow}_j + \ketbra{\downarrow}_j$\\
    $\ketbra{1}_j+\ketbra{-1}_j$ & $\ketbra{d}_j + \ketbra{h}_j$\\
    \colrule
    \multicolumn{2}{c}{\textbf{B. Operators $O$ with $[O,Q^\dagger] = Q^\dagger$, or $[O,\eta^\dagger] = \eta^\dagger$}}\\
    \colrule
    $\frac{1}{2}\sum_j S^z_j = Q^z$ & $\frac{1}{2}\sum_j \left(\ketbra{d}_j - \ketbra{h}_j \right) = \eta^z$ \\
    \botrule
    \end{tabular}
    \caption{On-site operators that either commute with $Q^\dagger$ or satisfy $[O, Q^\dagger] = Q^\dagger$. The two on-site operators in group A sum to identity.}
    \label{tab:table1}
\end{table}

We then focus our attention on two-site (nearest-neighbor) operators. The results in group A below hold on any bipartite lattice.  However, because we will rely on MPS techniques to find operators that annihilate $\ket{\mathcal{S}_N}$, we will simplify our notation to the 1D case going forward.

By elementary methods discussed in Appendix~\ref{sec:AppendixSystematicComm}, we find that the terms \#1-7 in Table~\ref{tab:operators} commute with $Q^\dagger$.

\begin{table*}
    \centering
    \begin{tabular}{c c c} \toprule
    \textbf{\#} & \textbf{Spin-1 model operator} & \textbf{Electronic model operator} \\
    \colrule
    \multicolumn{3}{c}{\textbf{A. Operators that commute with $Q^\dagger$, or $\eta^\dagger$} }\\
    \colrule
    \#1 & $\ketbra{00}_{j,j+1}$ & $\ketbra{\sigma_1, \sigma_2}{\sigma_3, \sigma_4}_{j,j+1} + \hc,~\sigma_1+\sigma_2 = \sigma_3 + \sigma_4$\\    
    \#2 & $(\ketbra{1}{1} + \ketbra{-1}{-1})_j(\ketbra{1}{1} + \ketbra{-1}{-1})_{j+1}$ & $(\ketbra{d}+\ketbra{h})_j(\ketbra{d}+\ketbra{h})_{j+1}$\\
    \#3 & $(\ketbra{1} + \ketbra{-1})_{j}(\ketbra{0})_{j+1}$ & $(\ketbra{d} +  \ketbra{h})_j (\ketbra{\uparrow} +  \ketbra{\downarrow})_{j+1}  $\\
    \#4 & $(\ketbra{0})_{j}(\ketbra{1} + \ketbra{-1})_{j+1}$ & $(\ketbra{\uparrow} +  \ketbra{\downarrow})_{j}(\ketbra{d} +  \ketbra{h})_{j+1}$\\
    \#5 & $e^{i\phi} \left(\ketbra{0,1}{1,0} - \ketbra{0,-1}{-1,0}  \right)_{j,j+1}+ \hc$ & ~~~$e^{i\phi} \big(\ketbra{\uparrow,d}{d,\uparrow} + \ketbra{\downarrow,d}{d,\downarrow}- \ketbra{\uparrow,h}{h,\uparrow} - \ketbra{\downarrow,h}{h,\downarrow}  \big)_{j,j+1}+ \hc$\\
    \#6 & $e^{i\phi} \ket{00}(\bra{1,-1}+\bra{-1,1})_{j,j+1} + \hc$ &$e^{i\phi}\left(\ket{\uparrow,\downarrow}-\ket{\downarrow,\uparrow} \right)\left(\bra{d,h}+\bra{h,d}\right)_{j,j+1} + \hc$ \\
    \#7 & $(\ket{1,-1}+\ket{-1,1})(\bra{1,-1}+\bra{-1,1})_{j,j+1}$ & $(\ket{d,h}+\ket{h,d})(\bra{d,h}+\bra{h,d})_{j,j+1}$ \\
    \colrule
    \multicolumn{3}{c}{\textbf{B. Operators $O$ with $[O,Q^\dagger] = Q^\dagger$, or $[O,\eta^\dagger] = \eta^\dagger$}}\\
    \colrule
    \#8 & $Q^z = \frac{1}{2}\sum_j S^z_j$ & $\eta^z = \frac{1}{2}\sum_j \left(\ketbra{d}_j - \ketbra{h}_j \right)$ \\
    \colrule
    \multicolumn{3}{c}{\textbf{C. Operators that annihilate scar tower} }\\
    \colrule
    \#9 & $\ketbra{\pm1,0}_{j,j+1}$ & $\ketbra{d}_{j}(\ketbra{\uparrow}+\ketbra{\downarrow})_{j+1}$  (or $\ketbra{h}_{j}$)\\
    \#10 & $\ketbra{0,\pm1}_{j,j+1}$ & $(\ketbra{\uparrow}+\ketbra{\downarrow})_{j}\ketbra{d}_{j+1}$ (or $\ketbra{h}_{j+1}$) \\
    \#11 & $e^{i\phi}\ketbra{0,\pm1}{\pm1,0}_{j,j+1} + \hc$ & $e^{i\phi}\left(\ketbra{\uparrow,d}{d,\uparrow}+\ketbra{\downarrow,d}{d,\downarrow}\right)_{j,j+1} + \hc$ (or $h$ instead of $d$)\\
    \#12 & $i\sum_j \left(\ketbra{-1,1}{1,-1}-\ketbra{1,-1}{-1,1}\right)_{j,j+1}$ & $i\sum_j \left(\ketbra{h,d}{d,h}-\ketbra{d,h}{h,d}\right)_{j,j+1}$\\
    \botrule
    \end{tabular}
    \caption{Nearest-neighbor operators that either commute with $Q^\dagger$, satisfy $[O,Q^\dagger] = Q^\dagger$, or annihilate the scar states $\ket{\mathcal{S}_N}$. In the left column, we show Hermitian operators in the spin-1 model that preserve $S^z_\text{Tot.}$, while in the right column we show translations to the electronic model in the ketbra notation (see Sec.~\ref{sec:translationtofermion}). Operators \#1-4 sum to the identity, and the on-site operators in Table~\ref{tab:table1} are linear combinations of \#1-4, e.g., $\ketbra{0}_j$ = \#1+\#4. We also note that in \#9-11, the $+1$ and $-1$ options are not independent---they are related by operators \#3-5 in group A. While the systematic procedure for finding the group C assumes translationally invariant operators on the 1D chain, we drop the summation over $j$ where it is not necessary: \#1-11 can be generalized to any bipartite lattice.}
    \label{tab:operators}
\end{table*}

We recover known terms that commute with $Q^\dagger$, such as $H_0$ in Eq.~(\ref{eq:H0}) (\#5 with $\phi=0$) and the pure biquadratic term $(\mathbf{S}_j \cdot \mathbf{S}_{j+1})^2$~\cite{moudgalya_large_2020} (\#1 $-$ \#6 ($\phi=0$) + \#7 + $I$). We lastly note that there are longer range terms that commute with $Q^\dagger$, such as $\sum_j \ketbra{0,0,0}_{j,j+1,j+2}$ --- one could also systematically construct them with such an approach.

\subsection{Operators with $[O, Q^\dagger] = Q^\dagger$ }
\label{subsec:opsthatcommutetoqdag}
The computation in the previous section also immediately gives us that the only nearest-neighbor operator that satisfies $[O, Q^\dagger] = Q^\dagger$ is in fact the on-site operator $Q^z = S^z_\text{Tot.}/2$ (\#8), modulo any linear combination of operators in group A (\#1-7 in Table~\ref{tab:operators}). 
We can prove this fact: given an $O$ such that $[O, Q^\dagger] = Q^\dagger$, we write $O = O - Q^z + Q^z$. It follows that $[O - Q^z, Q^\dagger] = 0$, so $O$ will be the sum of terms in group A and $Q^z$. Therefore $Q^z$ is the only independent term in group B.

\subsection{Operators that annihilate the scar tower}
\label{subsec:opsthatannihilate}
We next study nearest-neighbor operators that annihilate the scar tower $\ket{\mathcal{S}_N}$. To do so we compress all $\ket{\mathcal{S}_N}$ into a single Matrix Product State (MPS): $\ket{\mathcal{S}(z)} = \sum_N (z^N/N!) \ket{\mathcal{S}_N} = \exp(z Q^\dagger) \ket{\Omega}$, with parameter $z$. 
We then express nearest-neighbor operators as Matrix Product Operators (MPOs). Focusing on operators that conserve $S^z_\text{Tot.}$, it suffices to find all nearest-neighbor such operators that annihilate $\ket{\mathcal{S}(z)}$.  We adopt an approach similar to Ref.~\cite{chertkov_computational_2018}. In the MPO and MPS language this becomes a problem of finding all null eigenvectors of some small matrix [Eq.~(\ref{eq:annihilatingmatrix}) below]. The parameter $z$ is arbitrary, and for our numerical computations we take $z=0.1$. Before proceeding with details, we note that in the present case $\ket{\mathcal{S}(z)}$ is a product state, and we could in principle calculate without using the full MPO-MPS formalism (and even find all scarry models without compressing the scar tower, see Appendix~\ref{app:analyticSpin1Scarry}).  Nevertheless, the presented formalism is very powerful and can be applied as a black-box tool to any set of scars compressed into an MPS and will be applied also to the AKLT scar tower in Appendix~\ref{sec:AKLTfamily}.

Writing $\exp(zQ^\dagger)$ in MPO form:
\begin{gather}
    \exp(zQ^\dagger) = b_l^T \left(\prod_{j=1}^L M_j \right) b_r ~,\label{eq:MPO}\\
    M_j = \begin{pmatrix}
    -\mathbf{I}_j & -z(S^+_j)^2/2 \\
    z(S^+_j)^2/2 & \mathbf{I}_j
    \end{pmatrix},
    b_l = \begin{pmatrix}
    1 \\
    0
    \end{pmatrix},
    b_r = \begin{pmatrix}
    1 \\
    1
    \end{pmatrix}, \nonumber
\end{gather}
we can write $\ket{\mathcal{S}(z)}$ in MPS form:
\begin{gather}
    \ket{\mathcal{S}(z)} = \sum_{\{\sigma_j\}} b_l^T \left(\prod_{j=1}^L A^{[\sigma_j]} \right) b_r~ \ket{\{\sigma_j\}} ~, \label{eq:scarsMPS} \\
    A^{[-1]} = \begin{pmatrix}
    -1 & 0 \\
    0 & 1
    \end{pmatrix}, ~~
    A^{[0]} = \mathbf{0}, ~~
    A^{[1]} = z\begin{pmatrix}
    0 & -1 \\
    1 & 0
    \end{pmatrix} ~. \nonumber
\end{gather}
We express translationally invariant operators $O$ in the natural basis:
\begin{align}
    O &= \sum_{a+b=c+d} c_{ab,cd} \sum_j \ketbra{a,b}{c,d}_{j,j+1} \label{eq:translinvO} \\
    & = \sum_{a+b=c+d} c_{ab,cd} O_{ab,cd} ~.
\end{align}
We can express $O_{ab,cd}$ as an MPO:
\begin{gather}
O_{ab,cd} = (b_l^O)^T \left(\prod_{j=1}^{L} W_{ab,cd,j}\right) b_r^O~,\\
W_{ab,cd,j} = \left(\begin{array}{c|c} \begin{matrix}
  \mathbf{I}_j & \ketbra{a}{c}_j & 0 \\
  0 & 0  & \ketbra{b}{d}_j\\
  0 & 0 & \mathbf{I}_j
  \end{matrix} & \makebox(0,0){\text{\huge0}}\\
  \hline
  \makebox(0,0){\text{\huge0}}&\begin{matrix}
  0 & \ketbra{b}{d}_j & 0 \\
  0 & \mathbf{I}_j  & \ketbra{a}{c}_j\\
  0 & 0 & 0
  \end{matrix}\end{array}\right)~,\\
 b_l^O = \begin{pmatrix}
1 & 0 & 0 & 1 & 0 & 0
\end{pmatrix}^T~,\\
b_r^O = \begin{pmatrix}
0 & 0 & 1 & 0 & 0 & 1
\end{pmatrix}^T~.
\end{gather}
This lets us write $O_{ab,cd} \ket{\mathcal{S}(z)}$ in MPS form (an MPO$\times$MPS is an MPS, Ref.~\cite{moudgalya_entanglement_2018}):
\begin{gather}
    O_{ab,cd}\ket{\mathcal{S}(z)} = \sum_{\{\sigma_j\}} (\tilde{b}_l)^T \left(\prod_{j=1}^L B^{[\sigma_j]}_{ab,cd}\right) \tilde{b}_r~ \ket{\{\sigma_j\}}~,\\
    B^{[\sigma]}_{ab,cd} = \sum_{\sigma'} W^{\sigma,\sigma'}_{ab,cd} \otimes A^{[\sigma']}~,~ \tilde{b}_{l/r} = b_{l/r}^O \otimes b_{l/r}~. \nonumber
\end{gather}
Then for $O = \sum_{ab,cd} c_{ab,cd} O_{ab,cd}$ we evaluate $\ket{\psi} = O \ket{\mathcal{S}(z)}$. We want $\ket{\psi} = 0$, so it suffices to evaluate 
\begin{align}
    &\braket{\psi} =\\ &\sum_{\substack{a+b = c+d,\\a'+b' = c'+d'}} c^*_{a'b',c'd'}c_{ab,cd} \bra{\mathcal{S}(z)}O_{a'b',c'd'}^\dagger O_{ab,cd}^{\mathstrut} \ket{\mathcal{S}(z)}~. \nonumber
\end{align}
This becomes a linear problem in $c_{ab,cd}$, with matrix coefficients $\bra{\mathcal{S}(z)}O_{a'b',c'd'}^\dagger O_{ab,cd} \ket{\mathcal{S}(z)}$. We obtain this by forming the transfer matrix
\begin{equation}
    E_{a'b',c'd'; ab,cd} = \sum_{\sigma} B^{[\sigma]*}_{a'b',c'd'} \otimes B^{[\sigma]}_{ab,cd}
\end{equation}
and calculating
\begin{gather}
    \bra{\mathcal{S}(z)}O_{a'b',c'd'}^\dagger O_{ab,cd} \ket{\mathcal{S}(z)} = (b^E_l)^T E^L_{a'b',c'd';ab,cd} b_r^E~, \nonumber\\
    b^E_{l/r} = \tilde{b}_{l/r}^* \otimes \tilde{b}_{l/r}~.
    \label{eq:annihilatingmatrix}
\end{gather}
The desired coefficients $c_{ab,cd}$ are the null eigenvectors of the Hermitian matrix $\bra{\mathcal{S}(z)} O_{a'b',c'd'}^\dagger O_{ab,cd} \ket{\mathcal{S}(z)}$. In practice it suffices to use the case $L=16$, because our operators are strictly nearest-neighbor. (We have numerically verified that choosing larger $L$ does not give additional terms).

Given null eigenvectors $c_{ab,cd}$, we construct Hermitian operators by imposing the additional condition $c^*_{ab,cd} = c_{cd,ab}$. This procedure found the new operators \#9-12 in Table~\ref{tab:operators}. Although this procedure was done for translationally invariant operators, only for \#12 is translational invariance needed, while similar to the group A, terms \#9-11 can be applied on each bond independently.

\subsubsection{Null operators}
\label{subsubsec:nullops}
One must take some care to recognize ``null operators" in this nearest-neighbor operator basis, that is, operators that appear non-trivial but vanish under summation over $j$. For operators that conserve $S^z_\text{Tot.}$, there are two linearly independent null operators:
    \begin{multline}
        \sum_j \big(\ketbra{1,-1} + \ketbra{1,0} \\- \ketbra{-1,1} -\ketbra{0,1}\!\big)_{j,j+1}\\
        = \sum_j \left(\ketbra{1}_j - \ketbra{1}_{j+1} \right) = 0~,\label{eq:sum1j1j+1} 
    \end{multline}
    \vspace{-20pt}
    \begin{multline}
        \sum_j \big(\ketbra{0,-1} + \ketbra{0,1} \\- \ketbra{-1,0} - \ketbra{1,0} \big)_{j,j+1}\\
        = \sum_j \left(\ketbra{0}_j - \ketbra{0}_{j+1} \right) = 0~. \label{eq:sum0j0j+1}
    \end{multline}
The null operator corresponding to $\sum_j (\ketbra{-1}_j - \ketbra{-1}_{j+1})$ is linearly dependent in this basis to the above two terms.  The MPS procedure finds all operators of the form Eq.~(\ref{eq:translinvO}) that annihilate $\ket{\mathcal{S}(z)}$ but by itself does not know that some of them are null operators, and as such these two terms will have to be subtracted from the null space obtained by this procedure.

For example, the procedure above gives the seemingly new term $\sum_j \left(\ketbra{1,-1} - \ketbra{-1,1} \right)_{j,j+1}$ that annihilates $\ket{\mathcal{S}_N}$. This term also has an appealing interpretation of measuring the number of ``left" domain walls ``$1,-1$" minus the number of ``right" domain walls ``$-1,1$", which always cancel in $\ket{\mathcal{S}_N}$ because there are no 0s. However, by Eq.~(\ref{eq:sum1j1j+1}), this term is actually equivalent to $\sum_j \left(\ketbra{0,1} - \ketbra{1,0} \right)_{j,j+1}$, which is \#10$-$\#9 in Table~\ref{tab:operators}. We also note that of the four choices in \#9,10, only 3 are linearly independent, modulo Eq.~(\ref{eq:sum0j0j+1}).

For non--$S^z_\text{Tot.}$-preserving operators, specifically those that change $S^z_\text{Tot.}$ by 1 or 2, the null operators are $\sum_j e^{i\phi}(\ketbra{1}{0}_j - \ketbra{1}{0}_{j+1}) + \hc$ and $\sum_j e^{i\phi}(\ketbra{0}{-1}_j - \ketbra{0}{-1}_{j+1}) + \hc$, or $\sum_j e^{i\phi}(\ketbra{1}{-1}_j - \ketbra{1}{-1}_{j+1}) + \hc$ respectively.

It is possible to eliminate the null operators from the start, e.g., by representing range-$k$ translationally invariant operators in the basis of length-$k$ ``Gell-Mann strings" $\sum_j \lambda^{(\mu_1)}_j \cdots\lambda^{(\mu_k)}_{j+k-1}$ (spin-1 analog of ``Pauli strings" in spin-1/2 chains), where the start of the string $\lambda^{(\mu_1)}$ must be a non-trivial Gell-Mann matrix while the other positions $\lambda^{(\mu_i)},~2\leq i \leq k$ can also be an identity matrix. For range-2 operators, our overcomplete basis choice with proper treatment of the ``null operators" is equivalent but is more symmetric between the two sites and makes it somewhat easier to unpack results of the black-box search.  A similar use of two-site operator basis (overcomplete for range-2 translationally invariant operators) is also very convenient for the analytical argument in App.~\ref{app:analyticSpin1Scarry}.

\subsubsection{Discussion of results}
\label{sec:discususionofresults}
While it is easy to see that \#9-11 annihilate $\ket{\mathcal{S}_N}$ because $\ket{\mathcal{S}_N}$ contain no 0s, that \#12 does is less immediate. 

\#12 can be viewed as an exchange of ``$1,-1$" and ``$-1,1$", with opposite sign for the two directions of the process. We can see that it annihilates $\ket{\mathcal{S}_N}$ by a preimage argument: The image only contains product states of 1s and $-1$s. There are equal numbers of ``$1,-1$" and ``$-1,1$" strings. For each ``$\pm1,\mp1$", there is a preimage where that string was ``$\mp1,\pm1$". Each of these preimages has the same sign in $\ket{\mathcal{S}_N}$, and therefore they cancel out in the image. 

An alternate proof is the observation that
\begin{gather}
i \text{\#12} + \sum_j \left(\ketbra{1,-1}- \ketbra{-1,1}\right)_{j,j+1} \\
= \sum_j \left(\ket{1,-1}-\ket{-1,1}\right)\left(\bra{1,-1}+\bra{-1,1}\right)_{j,j+1}~, \nonumber\\
\implies i \text{\#12} =  \sum_j \Big(\left(\ket{1,-1}-\ket{-1,1}\right)\left(\bra{1,-1}+\bra{-1,1}\right)  \nonumber\\
+ \ketbra{1,0} - \ketbra{0,1}\Big)_{j,j+1}~, \label{eq:term12rewrite}
\end{gather}
with $\bra{1,-1} + \bra{-1,1}$, $\bra{1,0}$ and $\bra{0,1}$ annihilating $\ket{\mathcal{S}_N}$.  While \#12 can be rewritten into a sum of local terms that annihilate $\ket{\mathcal{S}_N}$, this is only possible after the non-local re-writing in Eq.~(\ref{eq:sum1j1j+1}). This non-local cancellation is captured by our MPS method.

We remark that the terms \#1-7, \#9-11 correspond to the $S^z_\text{Tot.}$-conserving part of the Shiraishi-Mori structure discussed in Refs.~\cite{schecter_weak_2019, mark_unified_2020}, while here we have additionally separated them into non-scarry (pseudospin-SU(2)-symmetric) and true scarry terms. The non-local term \#12 is new and does not appear to fit under this Shiraishi-Mori framework, at least with projectors in the space of nearest-neighbor operators.

We can show that the operator \#12 is indeed independent of this Shiraishi-Mori space as follows: Since \#12 is purely imaginary, it suffices to show that it is independent of the other purely imaginary terms. These are \#5,6, and 11, with $\phi=\pi/2$. Consider the state $\ket{1,1,...,1,-1_j,1,...,1}$. While \#5,11 annihilate it and \#6 can send it to either $i\ket{1,1,...,0,0_j,1,...,1}$ or $i\ket{1,1,...,0_j,0,1,...,1}$, \#12 maps this to $i\ket{1,1,...,1,-1_{j-1},1,...,1} - i\ket{1,1,...,1,-1_{j+1},1,...,1}$. Thus, \#12 indeed cannot be written as a superposition of \#5,6, and 11.

While the terms \#1-7 and \#9-11 generalize to any bipartite graph, with arbitrary bond-dependent couplings, the term \#12 appears special to the 1D chain with PBC and with one coupling.  However, from the nature of the scar states, we see that such a term placed on any closed loop of any bipartite lattice also annihilates the scar states, which allows us to generate analogs of \#12 on any such lattice with arbitrary loop-dependent couplings.
Equivalently, the following term
\begin{equation}
i \sum_{\expval{\mathbf{ij}}, \mathbf{i}\in A, \mathbf{j}\in B} f_{\mathbf{ij}} \left(\ketbra{-1,1}{1,-1} - \ketbra{1,-1}{-1,1} \right)_{\mathbf{i,j}} ~,
\end{equation}
with $\sum_\mathbf{i} f_\mathbf{ij} = 0$ for all $\mathbf{i}$ from sublattice $A$ and $\sum_\mathbf{j} f_\mathbf{ij} = 0$ for all $\mathbf{j}$ from sublattice $B$, annihilates the scar states.
As an example, if every site on the bipartite lattice has even degree~\cite{diestel2017basics}, we can achieve the above with $f_{\mathbf{ij}} = \pm 1$ by assigning an orientation to every bond such that on every site the number of inward bonds equals the number of outward bonds (a so-called ``Eulerian orientation").

As discussed in Appendix~\ref{sec:exhaustivesearch}, to find non--$S^z_\text{Tot.}$-preserving terms that annihilate $\ket{\mathcal{S}_N}$, it suffices to consider only terms that change $S^z_\text{Tot.}$ by a fixed amount. We find that the following terms annihilate $\ket{\mathcal{S}_N}$:
\begin{gather}
    (\alpha\ket{0,\pm1} + \beta\ket{\pm1,0})\bra{0,0}_{j,j+1} + \hc~,\\
    (\alpha\ket{0,\pm1} + \beta\ket{\pm1,0})(\bra{1,-1} + \bra{-1,1})_{j,j+1} + \hc~, \nonumber\\
    (\alpha\ket{0,1} + \beta\ket{1,0})(\gamma\bra{0,-1} + \delta\bra{-1,0})_{j,j+1} + \hc~, \nonumber
\end{gather}
for arbitrary complex $\alpha,\beta,\gamma,\delta$, after removing the ``null operators."
These terms correspond to the non--$S^z_\text{Tot.}$-preserving part of the Shiraishi-Mori structure in Refs.~\cite{schecter_weak_2019, mark_unified_2020}, which hence exhausts such nearest-neighbor scarry Hamiltonians.

In Appendix~\ref{app:analyticSpin1Scarry} we provide an analytic derivation of the results in this section. We use this derivation in Appendix~\ref{sec:q0towers} to study the $k=0$ bimagnon tower: $\ket{\mathcal{S}^{k=0}_N} = (Q^\dagger_{k=0})^N \ket{\Omega}$, where $Q^\dagger_{k=0} = \frac{1}{2}\sum_j (S_j^+)^2$. 
We further consider the related spin-1/2 $k=0$ magnon tower and find natural scarred models for both towers of states.

In Appendix~\ref{sec:exhaustivesearch} we prove that our systematic search is exhaustive in the space of nearest-neighbor models. In Appendix~\ref{sec:AKLTfamily} we then systematically construct all spin-1 nearest-neighbor operators that annihilate the AKLT scar tower~\cite{Moudgalya2018}, revealing that the only family of nearest-neighbor models containing the AKLT scar tower is
\begin{equation}
H_{AKLT} + h S^z_\text{Tot.} + \sum_j \!\sum_{m,n =-2}^0 \!\!\!\! c_{m,n}(j) \ketbra{T_{2,m}}{T_{2,n}}_{j,j+1} ~,
\end{equation}
where $\ket{T_{2,0}} =( \ket{1,-1} + 2\ket{0,0} + \ket{-1,1})/\sqrt{6}$, $\ket{T_{2,-1}} = (\ket{0,-1} + \ket{-1,0})/\sqrt{2}$, and $\ket{T_{2,-2}} = \ket{-1,-1}$, and $c_{nm}(j) = c_{mn}^*(j)$ for Hermiticity but can vary with $j$. This corroborates the results in Refs.~\cite{mark_unified_2020, moudgalya_large_2020}.

\section{Electronic models with eta-pairing scar tower from spin-1 models with XY tower}
\label{sec:translationtofermion}

\begin{table*}
    \centering
    \begin{tabular}{c c c} \toprule
    \textbf{\#} & \textbf{Ketbra notation} & \textbf{Operator notation} \\
    \colrule
    \multicolumn{3}{c}{  \textbf{A. Operators that commute with $\eta^\dagger$} }\\
    \colrule
    \#1 & $\ketbra{\sigma_1, \sigma_2}{\sigma_3, \sigma_4}_{j,j+1} + \hc,~\sigma_1+\sigma_2 = \sigma_3 + \sigma_4$ & $\mathbf{S}_j\cdot\mathbf{S}_{j+1}$\\
    && or $(n_{j,\up} + n_{j,\dn} - 2n_{j,\up} n_{j,\dn})(n_{j+1,\up} + n_{j+1,\dn} - 2n_{j+1,\up} n_{j+1,\dn})$\\
     \#2 & $(\ketbra{d}+\ketbra{h})_j(\ketbra{d}+\ketbra{h})_{j+1}$& $4\left(\left(n_{j,\uparrow}-\frac{1}{2}\right)\left(n_{j,\downarrow}-\frac{1}{2}\right)+\frac{1}{4}\right)\left(\left(n_{j+1,\uparrow}-\frac{1}{2}\right)\left(n_{j+1,\downarrow}-\frac{1}{2}\right)+\frac{1}{4}\right)$ \\
    \#3 & $(\ketbra{d} +  \ketbra{h})_j (\ketbra{\uparrow} +  \ketbra{\downarrow})_{j+1}  $ & $2\left(\left(n_{j,\uparrow}-\frac{1}{2}\right)\left(n_{j,\downarrow}-\frac{1}{2}\right)+\frac{1}{4}\right)\left(n_{j+1,\uparrow}P_{j+1,\downarrow}+ P_{j+1,\uparrow}n_{j+1,\downarrow} \right)$\\
    \#4 & $(\ketbra{\uparrow} +  \ketbra{\downarrow})_{j}(\ketbra{d} +  \ketbra{h})_{j+1}$ & $2\left(n_{j,\uparrow}P_{j,\downarrow}+ P_{j,\uparrow}n_{j,\downarrow} \right)\left(\left(n_{j+1,\uparrow}-\frac{1}{2}\right)\left(n_{j+1,\downarrow}-\frac{1}{2}\right)+\frac{1}{4}\right) $\\
    \#5 & $e^{i\phi} \big(\ketbra{\uparrow,d}{d,\uparrow} + \ketbra{\downarrow,d}{d,\downarrow}- \ketbra{\uparrow,h}{h,\uparrow}$& $-\sum_\sigma e^{i\phi}(c^\dagger_{j+1,\sigma} c^{\mathstrut}_{j,\sigma}n_{j,-\sigma}n_{j+1,-\sigma}+c^\dagger_{j,\sigma} c^{\mathstrut}_{j+1,\sigma}P_{j,-\sigma}P_{j+1,-\sigma}) + \hc$\\
    &\multicolumn{1}{c}{$ - \ketbra{\downarrow,h}{h,\downarrow}  \big)_{j,j+1}+ \hc$} \\
    \#6 &$e^{i\phi}\left(\ket{\uparrow,\downarrow}-\ket{\downarrow,\uparrow} \right)\left(\bra{d,h}+\bra{h,d}\right)_{j,j+1} + \hc$ & ~~~$\sum_{\sigma} e^{i\phi} \big(c^\dagger_{j+1,\sigma}c^{\mathstrut}_{j,\sigma} n_{j,-\sigma}P_{j+1,-\sigma} +c^\dagger_{j,\sigma}c^{\mathstrut}_{j+1,\sigma} n_{j+1,-\sigma}P_{j,-\sigma} \big)   + \hc$\\
    \#7 & $(\ket{d,h}+\ket{h,d})(\bra{d,h}+\bra{h,d})_{j,j+1}$ & 
    $n_{j,\up} n_{j,\dn} P_{j+1,\up} P_{j+1,\dn} + P_{j,\up} P_{j,\dn} n_{j+1,\up} n_{j+1,\dn} + (c^\dagger_{j,\uparrow}c^\dagger_{j,\downarrow}c^{\mathstrut}_{j+1,\downarrow}c^{\mathstrut}_{j+1,\uparrow} + \hc)$\\
    \colrule
    \multicolumn{3}{c}{\textbf{B. Operators with $[O, \eta^\dagger] = \eta^\dagger$}}\\
    \colrule
    \#8 &   $\frac{1}{2}\left(\ketbra{d}_j - \ketbra{h}_j \right)$
    & $\frac{1}{2}\sum_{j,\sigma} (n_{j,\sigma} - \frac{1}{2})$\\
    \colrule
    \multicolumn{3}{c}{  \textbf{C. Operators that annihilate scar tower} }\\
    \colrule
    \#9 & $\ketbra{d}_{j}(\ketbra{\uparrow}+\ketbra{\downarrow})_{j+1}$  (or $\ketbra{h}_{j}$)& $ n_{j,\uparrow}n_{j,\downarrow} \left(n_{j+1,\uparrow}P_{j+1,\downarrow}+ P_{j+1,\uparrow}n_{j+1,\downarrow} \right)$ (or $P_{j,\uparrow}P_{j,\downarrow}$ instead of $n_{j,\uparrow}n_{j,\downarrow}$) \\
    \#10 & $(\ketbra{\uparrow}+\ketbra{\downarrow})_{j}\ketbra{d}_{j+1}$ (or $\ketbra{h}_{j+1}$) & $ \left(n_{j,\uparrow}P_{j,\downarrow}+ P_{j,\uparrow}n_{j,\downarrow} \right)n_{j+1,\uparrow}n_{j+1,\downarrow}$ (or $P_{j+1,\uparrow}P_{j+1,\downarrow}$ instead of $n_{j+1,\uparrow}n_{j+1,\downarrow}$)\\
    \#11 & $e^{i\phi}\left(\ketbra{\uparrow,d}{d,\uparrow}+\ketbra{\downarrow,d}{d,\downarrow}\right)_{j,j+1} + \hc$  & $-\sum_{\sigma} e^{i\phi} c^\dagger_{j+1,\sigma}c^{\mathstrut}_{j,\sigma}n_{j,-\sigma}n_{j+1,-\sigma}+ \hc $\\
    &(or $h$ instead of $d$)&  (or $\sum_{\sigma} e^{i\phi} c^\dagger_{j,\sigma}c^{\mathstrut}_{j+1,\sigma}P_{j,-\sigma}P_{j+1,-\sigma}+ \hc $)\\
    \#12 & $i\sum_j \left(\ketbra{h,d}{d,h}-\ketbra{d,h}{h,d}\right)_{j,j+1}$ & $\sum_j \left(i c^\dagger_{j+1,\uparrow} c^\dagger_{j+1,\downarrow} c^{\mathstrut}_{j,\downarrow} c^{\mathstrut}_{j,\uparrow} + \hc \right)$\\
    \botrule
    \end{tabular}
    \caption{Translation of electronic operators from ketbra notation to operator notation.  Here we use the projector $P_{j,\sigma}$ as shorthand for $1-n_{j,\sigma}$.
    }
    \label{tab:operatorstranslate}
\end{table*}

As discussed in Sec.~\ref{sec:relationbetweenfermionandspin}, the similarity between the electronic eta-pairing operator $\eta^\dagger$ and the spin-1 operator $Q^\dagger$ suggests making the identification $\ket{-1} \rightarrow \ket{h}, \ket{1} \rightarrow \ket{d}$. We can identify $\ket{0}$ either with $\ket{\uparrow}$ or $\ket{\downarrow}$. This identification is useful to convert our results in Sec.~\ref{sec:spinmodels} into the electronic setting.  By making the natural choice to restrict to electronic models with spin-SU(2) invariant operators (which combined with electron number conservation imply separate conservation of both $\up$ and $\dn$ species), we find that almost every
spin-1 model operator has a unique mapping to an electronic model operator.

\subsection{Ketbra notation in electronic models}

Before we discuss the procedure we make a comment on notation. While in the spin setting the notation $\ketbra{a,b}{c,d}$ is unambiguous, in the electronic setting some care has to be taken. The notation $\ketbra{a,b}{c,d}$ is well defined when we restrict our discussion to nearest-neighbor terms in 1D (and naturally to terms that conserve fermion number parity).

We adopt the Fock space ordering convention that the $c^\dagger$s should be ordered first with larger site numbers to the right, then with $\downarrow$s to the right of $\uparrow$s. Therefore we identify the kets:
\begin{equation*}
    \ket{\{n_{j,\alpha} \}} = (c_{1,\up}^\dagger)^{n_{1,\up}} (c_{1,\dn}^\dagger)^{n_{1,\dn}} (c_{2,\up}^\dagger)^{n_{2,\up}} (c_{2,\dn}^\dagger)^{n_{2,\dn}} \dots \ket{\text{vac.}}. 
\end{equation*}
We adopt the notation that (for the part of the ket associated with site $j$) $\ket{h_j} = \ket{n_{j,\up}=0, n_{j,\dn}=0}$, 
$\ket{d_j} = \ket{n_{j,\up}=1, n_{j,\dn}=1}$, $\ket{\up_j} = \ket{n_{j,\up}=1, n_{j,\dn}=0}$, and $\ket{\dn_j} = \ket{n_{j,\up}=0, n_{j,\dn}=1}$.

Diagonal single-site operators are independent of the choice of convention:
\begin{align}
\ketbra{h}_j &\leftrightarrow (1 - n_{j,\up}) (1 - n_{j,\dn}) ~, \\  
\ketbra{d}_j &\leftrightarrow n_{j,\up} n_{j,\dn} ~, \nonumber\\
\ketbra{\up}_j &\leftrightarrow n_{j,\up} (1 - n_{j,\dn}) ~, \nonumber
\end{align}
and likewise for $\ketbra{\dn}_j$. However, our convention allows the ketbra notation to be sensible for single-site off-diagonal terms and nearest-neighbor terms with even fermion parity. For example,
\begin{align}
   \ketbra{d}{h}_j \leftrightarrow c_{j\up}^\dagger c_{j\dn}^\dagger ~, \label{eq:splus}\\ 
   \ketbra{\up}{\dn}_j \leftrightarrow c_{j,\up}^\dagger c_{j,\dn}^{\mathstrut} ~, \nonumber
\end{align}
etc. In particular, we have a direct connection between the $Q^\dagger$ and $\eta^\dagger$ operators, including all signs. However, terms like $\ketbra{\up}{h}_j$ are not well-defined (i.e., cannot appear in the electronic Hamiltonian by themselves), because they change total fermion parity. Hence it is natural to only consider operators where this does not happen.

We next consider two-site operators. We restrict ourselves to nearest-neighbor operators, because it is clear that in general, operators $f_j g_k$ composed of fermionic operators $f_j$ and $g_k$ depend on the occupations between sites $j$ and $k$ and do not have well defined (localized) ketbra representations. Our Fock space convention allows unambiguous conversion between operator and ketbra notation for nearest neigbbor terms. For example, consider the operator $c^\dagger_{j,\uparrow} c_{j+1,\uparrow}^{\mathstrut} n_{j,\downarrow} n_{j+1,\downarrow}$, which sends $\ket{\dots \downarrow_j, d_{j+1} \dots} \rightarrow \pm \ket{\dots d_j, \downarrow_{j+1} \dots}$. The corresponding sign of $\ketbra{d,\downarrow}{\downarrow,d}_{j,j+1}$ is determined by writing:
\begin{align}
    c^\dagger_{j,\uparrow} c^{\mathstrut}_{j+1,\uparrow} &n_{j,\downarrow} n_{j+1,\downarrow} \ket{\downarrow,d}_{j,j+1} \\
    &= c^\dagger_{j,\uparrow} c^{\mathstrut}_{j+1,\uparrow} \left(c^\dagger_{j,\downarrow} c^\dagger_{j+1,\uparrow} c^\dagger_{j+1,\downarrow} \ket{\text{vac.}} \right) \nonumber\\
    &= -c^\dagger_{j,\uparrow} c^\dagger_{j,\downarrow} c^\dagger_{j+1,\downarrow} \ket{\text{vac.}} = -\ket{d,\downarrow}_{j,j+1} ~. \nonumber
\end{align}
Therefore $c^\dagger_{j,\uparrow} c_{j+1,\uparrow}^{\mathstrut} n_{j,\downarrow} n_{j+1,\downarrow} \leftrightarrow -\ketbra{d,\downarrow}{\downarrow,d}_{j,j+1}$.~\footnote{For an electronic chain with periodic boundary conditions, terms that contain single electron hopping across the $(L,1)$ bond acquire an additional sign $(-1)^{N_\text{el.tot.} - 1}$ when going between the electron operator and ketbra expressions compared to $(j,j+1)$ terms with $j<L$.}

The ketbra notation is not well defined (does not give nice localized expressions) in every Fock space convention. Consider the convention where all up spins are listed before down spins, then site-ordered within each spin grouping. This convention, while convenient in numerical computations, is not a local one and thus does not support the nice ketbra notation.

\subsection{Mapping spin-1 model operators to electronic model operators}
Having clarified the meaning of $\ketbra{a,b}{c,d}$ for fermions, we can map spin-1 model operators to electronic model operators. In each case we map $1\rightarrow d~, -1\rightarrow h$, but the $0$ term requires some thought. 
There are several cases:
\begin{itemize}
    \item Only 1s and -1s appear in the ket and bra. Here the mapping is straightforward, and $d$ behaves as a boson hopping over $h$. \#2,7, and 12 fall in this case.
    \item One 0 appears in the ket and bra each. To conserve each fermion spin species, we map the 0 in both the bra and ket to the same species.
    \#3,4,5,9,10,11 fall in this case, and we map, for example, $\ketbra{0,1}\rightarrow \ketbra{\uparrow,d}$ or $\ketbra{\downarrow,d}$.
    \item Two zeros appear in either the ket or the bra, e.g., $\ketbra{00}{1,-1}$. To conserve the fermion numbers, $\ket{00}\bra{1,-1}\rightarrow \ket{\uparrow,\downarrow}\bra{d,h}$ or $\ket{\downarrow,\uparrow}\bra{d,h}$. \#6 belongs to this case.
    \item $\ketbra{00}{00}$. Any `00' mappings that conserve the fermion numbers are allowed, e.g., $\ketbra{\uparrow\uparrow},\ketbra{\uparrow\downarrow}{\downarrow,\uparrow}$, etc. \#1 belongs to this case.
\end{itemize}

Therefore we see that by just requiring separate conservation of $\up$ and $\dn$ electron species, each term has at most two possible mappings (except the $\ketbra{00}{00}$ term). Further requiring spin-SU(2) invariance of the electronic terms leads to a unique choice of mapping, e.g., $\ketbra{0,1}{1,0} \rightarrow \ketbra{\uparrow,d}{d,\uparrow} + \ketbra{\downarrow,d}{d,\downarrow}$ and $(\ket{1,-1}+\ket{-1,1})\bra{0,0} \rightarrow (\ket{d,h}+\ket{h,d})(\bra{\uparrow,\downarrow}-\bra{\downarrow,\uparrow})$. One can see the required signs most easily in the ketbra notation, because we do not have to worry about signs when applying the electronic $S^\pm_j$ operators onto kets, see Eq.~(\ref{eq:splus}). For example, for \#6:
\begin{align}
    &\left[S^+_{\text{Tot}}, e^{i\phi} (\ket{\uparrow,\downarrow} + \gamma \ket{\downarrow,\uparrow})(\bra{d,h}+\bra{h,d}) + \hc \right] \\
    &= e^{i\phi}(1 + \gamma)\ket{\uparrow,\uparrow}(\bra{d,h}+\bra{h,d}) \nonumber\\
    &~~~~-e^{-i\phi}(1+\gamma^*)(\ket{d,h}+\ket{h,d})\bra{\downarrow,\downarrow}~.\nonumber
\end{align} 
This commutator is only zero when $\gamma = -1$, fixing the sign of the spin-SU(2) invariant mapping.

The case of $\ketbra{00}$ is an exception. In this case there are two spin-SU(2) symmetric terms allowed: $\mathbf{S}_j \cdot \mathbf{S}_{j+1}$ and $\sum_{\sigma,\sigma'}\ketbra{\sigma,\sigma'}_{j,j+1} =
(n_{j,\up} + n_{j,\dn} - 2n_{j,\up} n_{j,\dn})(n_{j+1,\up} + n_{j+1,\dn} - 2n_{j+1,\up} n_{j+1,\dn})$
(projector onto the space of no holons or doublons).

This allows us to complete the right column in Table~\ref{tab:operators}. By applying the procedure discussed above, we translate these operators in terms of standard electronic operators in Table~\ref{tab:operatorstranslate}. 
A subtle point is that the terms \#5,6, and \#11 corresponding to single electron hopping cannot be immediately mapped in their translationally invariant (i.e., summed over $j$) form, because there are sign ambiguities with hopping between sites $1$ and $L$. However, because we can specialize to bond-wise operators satisfying the desired commutation or scar annihilation properties, the mapping can proceed, by suitable redefining of the site numbers, or ordering convention, for hopping across the ends (for more details, see Sec.~\ref{subsec:gentobip} on generalizing these results to arbitrary bipartite lattices). Only \#12 needs to be mapped in its translationally invariant form. Here the conversion is valid in PBC because the term moves two electrons and the simple-minded ketbra notation still holds across the endpoints.

While we verified directly in the fermion picture that the electronic terms have the desired properties with $\eta^\dagger$ or $\ket{\psi_N}$, these properties are in fact immediate from the spin-1 to electronic model translation. 
When considering group A, $Q^\dagger$ only involves $\ket{\pm 1}$ while $\eta^\dagger$ only involves $\ket{d},\,\ket{h}$. Except for the case of $\ketbra{00}$, due to the electronic spin-SU(2) constraint, operators involving the ``irrelevant" state $\ket{0}$ are mapped uniquely onto electronic operators involving $\ket{\up},\,\ket{\dn}$.
Therefore we can establish a bijection between spin-1 model and electronic model nearest-neighbor operators, and the commutation result carries from one setting to the other.

Likewise, for group C, terms \#9-11 have $\up/\dn$ in the bra's of the ketbra and trivially annihilate the eta-pairing scar tower. \#12 involves only $h$ and $d$, and the results from the spin-1 setting carry over with no issue.

A point of curiosity is that the analog of the minus sign in the spin-1 model $H_0$ (Eq.~(\ref{eq:H0}), \#5 with $\phi = 0$) is naturally present in the electronic hopping terms with real-valued hopping amplitude. The relative minus sign is important for $H_0$ to be pseudospin SU(2) symmetric [and for the electronic model that it maps to to be eta-pairing SU(2) symmetric]. In the spin-1 setting for physical spins it is less natural to have a sign difference between $\ketbra{0,-1}{-1,0}$ and $\ketbra{0,1}{1,0}$ spin exchange processes, and so the ``natural'' spin-1 XY model studied by Schecter and Iadecola breaks the pseudospin symmetry.

In electronic models, however, due to fermionic anticommutation, the electronic hopping Hamiltonian (\#5) gives opposite signs to the terms $\ketbra{\sigma,h}{h,\sigma}$ and $\ketbra{\sigma,d}{d,\sigma}$.  We recover the pure hopping term from \#6$-$\#5 in Table~\ref{tab:operatorstranslate} (with $\phi=0$ in both terms), in agreement with the fact that the electronic hopping, and thus the Hubbard model, is eta-pairing SU(2) symmetric.

To address our original goal of finding simplest models that turn the eta-pairing states into true scar states, we additionally require that the operators respect inversion and time-reversal symmetry. This leaves us with the combinations of \#9 and \#10 --- $\sum_j (\ketbra{\uparrow} + \ketbra{\downarrow})_j \ketbra{d}_{j+1} + \ketbra{d}_{j} (\ketbra{\uparrow} + \ketbra{\downarrow})_{j+1}$ and its $h$ equivalent --- and \#11. 
One would expect the first term to come from a Coulomb interaction $\frac{1}{2}\sum_j(n_{j,\up} + n_{j,\dn})(n_{j+1,\up} + n_{j+1,\dn})$. The other contributions to the density-density term are $\frac{1}{2}\sum_{j,\sigma,\sigma'}\ketbra{\sigma,\sigma'}{\sigma,\sigma'}_{j,j+1}$ and $2\sum_{j}\ketbra{d,d}{d,d}_{j,j+1}$. The former is a possible translation of \#1, but the latter is not in Table~\ref{tab:operatorstranslate}. Therefore the Coulomb interaction does not preserve the eta-pairing states as scars and we do not expect the first term to arise ``naturally."

The term \#11, however, is contained in the Hirsch term~[Eq.~(\ref{eq:Hirschmodel})]. Specifically, from Table~\ref{tab:operatorstranslate}, $H_\text{Hirsch} = X(\text{\#6} - 2\text{\#11})$, with $\phi=0$. We therefore conclude that not only is the Hirsch model scarred, it is one of few models to have the eta-pairing states as scars. The Hirsch model is also arguably the most ``natural" model in this family of scarred models, having been previously explored for different reasons.

We lastly note that by adding \#1, the eta-pairing scars are robust to the presence of spin-spin interactions. 

The procedure outlined in this work gives a framework for listing families of models which share an exact scar tower. An interesting question for future study might be to see if one can then similarly systematically identify sets of states as candidate scar towers, on which we can apply this method.

\subsection{Generalization to arbitrary bipartite lattice}
\label{subsec:gentobip}
While the procedure in Sec.~\ref{sec:spinmodels} assumed translational invariance, the individual terms in \#1-7 and \#9-11 across each bond $j,j+1$ satisfy the desired commutation or scar tower annihilation properties. In the spin-1 model setting, we can generalize these terms to any bond $\expval{\mathbf{ij}}$ between sites $\mathbf{i}$ and $\mathbf{j}$ on different bipartite components. This also turns out to be true in the electronic setting. There is no Fock space ordering convention that simultaneously preserves locality (in the sense discussed in this section) for every bond $\expval{\mathbf{i,j}}$. 
However, for a given bond $\expval{\mathbf{ij}}$, we can define a convention that orders site $\mathbf{j}$ operators immediately after site $\mathbf{i}$ operators, for example. We can then define the ketbra $\ketbra{...}{...}_\mathbf{i,j}$ in this convention and prove the commutation or scar annihilation properties for the operator on this bond.

Lastly, as discussed in Sec.~\ref{sec:discususionofresults}, we can generalize \#12 onto bipartite lattices by writing it as an operator on closed loops in the graph.

\section{Conclusion}
We observed that in the Hirsch model, the eta-pairing states --- exactly known states in the Hubbard model --- become many-body scar states, because the eta-pairing SU(2) symmetry is broken. While this observation is easy to verify, we arrived at this conclusion by first studying nearest-neighbor spin-1 models that are scarred by the spin-1 XY scar tower. Our systematic study separated models in which the spin-1 XY scar tower states are true scars from models that preserve the pseudospin SU(2) symmetry. We also found a new model (\#12) that lies outisde the established Shiraishi-Mori projector structure in the spin-1 XY model. These results give insights on other spin models with scar towers (Appendix~\ref{sec:q0towers}).

We then translated our findings from the spin-1 setting to the electronic setting, and obtained a family of models that are scarred by the eta-pairing states, of which the Hirsch model is a member. This work thus furthers our understanding of exact many-body scar towers and provides a systematic way of constructing families of scarred models.

{\it Note added}: While we were preparing our manuscript, we learned about related work by S.~Moudgalya, N.~Regnault, and B.~A.~Bernevig~\cite{moudgalya_eta-pairing_2020}, which will appear in the same arXiv posting.

\begin{acknowledgments}
We thank B.~Andrei Bernevig, Anushya Chandran, Matthew Fisher, Jim Garrison, Vedika Khemani, Ryan Mishmash, Sanjay Moudgalya, and Nicolas Regnault for valuable discussions, and in particular Cheng-Ju Lin for many discussions and collaborations on related topics. D.~K.~M.\ acknowledges funding from the James C.~Whitney SURF Fellowship, Caltech Student-Faculty Programs. This work was also supported by National Science Foundation through Grant DMR-1619696.

\end{acknowledgments}

\appendix

\section{Additional scar towers in the Hirsch model}
\label{sec:AppendixTriplSinglScars}
In addition to the eta-pairing states $\ket{\psi_N}$ in Eq.~(\ref{eq:eta pairing states}), we also find several more towers of scar states in the 1D Hirsch model [Eq.~(\ref{eq:Hirschmodel})]. We define these as:
\begin{align}
    \ket{\psi_{N,M}} &= (t^\dagger)^M \ket{\psi_N} = (t^\dagger)^M (\eta^\dagger)^N \ket{\text{vac.}},\\
    \ket{\phi_N} &= s^\dagger \ket{\psi_N} = s^\dagger (\eta^\dagger)^N \ket{\text{vac.}},
\end{align}
where
\begin{align}
  t^\dagger &= \sum_j (-1)^j c^\dagger_{j,\uparrow} c^\dagger_{j+1,\uparrow} ~,\\
  s^\dagger &= \sum_j (-1)^j \left(c^\dagger_{j,\uparrow} c^\dagger_{j+1,\downarrow} - c^\dagger_{j,\downarrow} c^\dagger_{j+1,\uparrow} \right) ~.
\end{align}
$t^\dagger$ creates a nearest-neighbor triplet $\ket{\uparrow, \uparrow}$ with momentum $\pi$, while $s^\dagger$ creates a nearest-neighbor singlet $\ket{\uparrow, \downarrow} - \ket{\downarrow, \uparrow}$ with momentum $\pi$. While $t^\dagger$ can be applied an arbitrary number of times to the eta-pairing states $\ket{\psi_N} = \ket{\psi_{N,0}}$, $s^\dagger$ can only be applied once to obtain the exact eigenstate $\ket{\phi_N} = s^\dagger \ket{\psi_N}$.

$\ket{\psi_{N,M}}$ and $\ket{\phi_N}$ both have energies $E = N U$, independent of the strength of the Hirsch term $X$. (Here and below, we set $\mu=0$ without loss of generality.) Additionally, $\psi_{N,M}$ has total spin $S=M$, momentum $k=(N+M)\pi~(\text{mod }2\pi)$ and site inversion number $I_s = 1$, while $\phi_N$ has spin $S=0$, momentum $k=(N+1)\pi~(\text{mod }2\pi)$ and site inversion number $I_s = -1$.

We prove that $\ket{\psi_{N,M}}$ and $\ket{\phi_N}$ are exact eigenstates using a commutator argument~\cite{mark_unified_2020}. We first note that the states of interest all have exactly $N$ doublons. Hence, it suffices to show that they are eigenstates---in fact annihilated by---the kinetic terms. In fact, since these states are independent of $X$, we show that each hopping process $H_{\text{kin},a}, a = 1,2,3$ independently annihilates the scar towers, where:
\begin{align}
    H_\text{kin,1} &= \sum_{j,\sigma} \left( \ketbra{h,\sigma}{\sigma,h}_{j,j+1} + \hc \right) ~,\\
    H_\text{kin,2} &= \sum_j \left[ \left(\ket{d,h} + \ket{h,d} \right) \left(\bra{\uparrow,\downarrow} - \bra{\downarrow,\uparrow} \right)_{j,j+1} + \hc \right] ~,\nonumber\\
    H_\text{kin,3} &= \sum_{j,\sigma} \left( \ketbra{d,\sigma}{\sigma,d}_{j,j+1} + \hc \right) ~.\nonumber
\end{align}
The expressions in terms of electron operators are obtained using the prescription described in Sec.~\ref{sec:translationtofermion} and Table~\ref{tab:operatorstranslate} ($H_\text{kin,1/3} \leftrightarrow$ \#11 and $H_\text{kin,2} \leftrightarrow$ \#6).  We take these electronic expressions with periodic boundary conditions, $c_{L+1,\sigma} \equiv c_{1,\sigma}$, as defining our model, but we use the above ketbra writings in terms of $h,\up,\dn,d$ as being more compact and better revealing structure in the arguments below. There is a small subtlety regarding hopping across the bond $(L,1)$~\footnotemark[\value{footnote}].
Since our states $\ket{\psi_{N,M}}$ and $\ket{\phi_N}$ always have an even number of electrons, the hopping terms across $(L,1)$ for $H_{\text{kin},a}$ flip sign. For example, the $H_\text{kin,1}$ term across $(L,1)$ is $-(\ketbra{h,\sigma}{\sigma,h}_{L,1} + \hc)$. While in the rest of our proofs we do not explicitly exhibit this subtlety, it is easy to verify that appropriate ketbra expressions for the operators $t^\dagger$ and $s^\dagger$ also have this flipped sign when the $\sigma,\sigma'$ pair is created across the bond $(L,1)$. These sign differences cancel out and our proofs remain valid.

The Hirsch Hamiltonian in Eq.~(\ref{eq:Hirschmodel}) can be written as $H = -t H_\text{kin,1} - (t-X) H_\text{kin,2} + (t-2X) H_\text{kin,3} + U \sum_j \ketbra{d}_j$. Given the fact that $\ket{\psi_{0,M}}$ and $\ket{\phi_0}$ are annihilated by $H_{\text{kin},a}$ (which we will prove below), if we show that $[H_{\text{kin},a}, \eta^\dagger] \ket{\psi_{N,M}} = [H_{\text{kin},a}, \eta^\dagger] \ket{\phi_N} = 0$, it follows that these states are annihilated by $H_{\text{kin},a}$.

From Table~\ref{tab:operatorstranslate} we know that $\left[H_\text{kin,2}, \eta^\dagger \right] = 0$ and $\left[H_\text{kin,1} - H_\text{kin,3}, \eta^\dagger \right] = 0$.  Then it suffices to show that $\left[H_\text{kin,1}, \eta^\dagger \right] \ket{\psi_{N,M}} = \left[H_\text{kin,1}, \eta^\dagger \right] \ket{\phi_{N}} = 0$.

Evaluating the commutator gives:
\begin{align}
    & \left[H_\text{kin,1}, \eta^\dagger \right] = \sum_{j,\sigma} \left[ \ketbra{h,\sigma}{\sigma,h}_{j,j+1} + \hc, \eta^\dagger \right] \\
    &= \sum_{j,\sigma} (-1)^j \left[ \ketbra{h,\sigma}{\sigma,h}_{j,j+1} + \hc, \ketbra{d}{h}_j - \ketbra{d}{h}_{j+1} \right] \nonumber\\
    & = \sum_{j,\sigma} (-1)^j \left(\ketbra{\sigma,d}{h,\sigma} - \ketbra{d,\sigma}{\sigma,h}\right)_{j,j+1} ~. \nonumber
\end{align}
This annihilates $\ket{\psi_{N,M}}$ and $\ket{\phi_N}$ because we note that in these states the spins appear in bound pairs, as elaborated below:

In $\ket{\psi_{N,M}}$, there are only $\uparrow$ unpaired electrons, which appear in nearest-neighbor pairs $\uparrow,\uparrow$. Then
\begin{align}
& \sum_\sigma (\ketbra{\sigma,d}{h,\sigma} - \ketbra{d,\sigma}{\sigma,h})_{j,j+1} \ket{\psi_{N,M}} = \\
&\left(\ketbra{\uparrow,d,\uparrow}{h,\uparrow,\uparrow}_{j,j+1,j+2} \! - \ketbra{\uparrow,d,\uparrow}{\uparrow,\uparrow,h}_{j-1,j,j+1} \right)\!\ket{\psi_{N,M}} .\nonumber
\end{align}
Summing over $j$, we get for $\left[H_\text{kin,1}, \eta^\dagger \right] \ket{\psi_{N,M}}$
\begin{align}
    &-\sum_j (-1)^{j} \ket{\uparrow,d,\uparrow}\left(\bra{h,\uparrow,\uparrow} + \bra{\uparrow,\uparrow,h} \right)_{j-1,j,j+1} \ket{\psi_{N,M}} \nonumber\\
    &= 0~.
\end{align}
The last line follows because the $\uparrow,\uparrow$ pair created by $t^\dagger$ has momentum $k=\pi$, and so such pairs appear in $\ket{\psi_{N,M}}$ in the superposition $\ket{h,\uparrow,\uparrow} - \ket{\uparrow,\uparrow,h}$.

In $\ket{\phi_n}$, there is one pair of nearest-neighbor $\uparrow$ and $\downarrow$ electrons in a singlet state. Then
\begin{align}
& \sum_{\sigma}(\ketbra{\sigma,d}{h,\sigma} - \ketbra{d,\sigma}{\sigma,h})_{j,j+1} \ket{\phi_N} \\
& = \big[(\ketbra{\uparrow,d,\downarrow}{h,\uparrow,\downarrow} + \ketbra{\downarrow,d,\uparrow}{h,\downarrow,\uparrow})_{j,j+1,j+2} \nonumber\\
&- (\ketbra{\downarrow,d,\uparrow}{\downarrow,\uparrow,h} + \ketbra{\uparrow,d,\downarrow}{\uparrow,\downarrow,h})_{j-1,j,j+1} \big] \ket{\phi_{N}} \nonumber~.
\end{align}
Summing over $j$ gives for $\left[H_\text{kin,1}, \eta^\dagger \right] \ket{\phi_N}$
\begin{align}
    &-\sum_j (-1)^j \Big[\ket{\uparrow,d,\downarrow}\left(\bra{h,\uparrow,\downarrow}+\bra{\uparrow,\downarrow,h}\right)_{j-1,j,j+1}  \nonumber\\
    &+\ket{\downarrow,d,\uparrow}\left(\bra{h,\downarrow,\uparrow}+\bra{\downarrow,\uparrow,h}\right)_{j-1,j,j+1} \Big] \ket{\phi_N} = 0~.
\end{align}
This is zero because the $\uparrow,\downarrow$ and $\downarrow,\uparrow$ pairs have momentum $\pi$.

We lastly have to show that the initial states $\ket{\psi_{0,M}}$ and $\ket{\phi_0}$ are annihilated by each $H_{\text{kin},a}$. The case of $\ket{\psi_{0,M}} = (t^\dagger)^M \ket{\text{vac.}}$ is immediate. The only nontrivial term is $H_\text{kin,1}$. Since the bound $\uparrow,\uparrow$ pairs in $\ket{\psi_{0,M}}$ can be constructed by creating electrons in momentum states of the form $c^\dagger_{\uparrow, k} c^\dagger_{\uparrow, \pi-k}$, whose kinetic energy is $\cos(k) + \cos(\pi-k) = 0$, we conclude that $\ket{\psi_{0,M}}$ has zero energy under $H_\text{kin,1}$.

In the case of $\ket{\phi_0} = s^\dagger \ket{\text{vac.}}$, we have to consider both $H_\text{kin,1}$ and $H_\text{kin,2}$. To show $H_\text{kin,1} \ket{\phi_0} = 0$, we write
\begin{align}
    &H_\text{kin,1} \sum_j (-1)^j (\ket{...,h,\uparrow_j,\downarrow,h,...} - \ket{...,h,\downarrow_j,\uparrow,h,...})\nonumber\\
    & = \sum_j (-1)^j \Big[\ket{...,\uparrow,h_j,\downarrow,h,...} - \ket{...,\downarrow,h_j,\uparrow,h,...} \nonumber\\
    & + \ket{...,h,\uparrow_j,h,\downarrow,...} - \ket{...,h,\downarrow_j,h,\uparrow,...}\Big] = 0~.
\end{align}
Lastly, $H_\text{kin,2} \ket{\phi_0} = 0$ because
\begin{align}
    &H_\text{kin,2} \sum_j (-1)^j (\ket{...,h,\uparrow_j,\downarrow,h,...} - \ket{...,h,\downarrow_j,\uparrow,h,...})\nonumber\\
    & = 2\sum_j (-1)^j \Big(\ket{...,h,d_j,h,h,...}+\ket{...,h,h_j,d,h,...}\Big) \nonumber\\
    & = 0~.
\end{align}
Note that the arguments for $[H_{\text{kin},1}, \eta^\dagger] \ket{\phi_N}= 0$ and $H_{\text{kin},a} \ket{\phi_0} = 0$ held for the $\ket{\uparrow,\downarrow}$ and $\ket{\downarrow,\uparrow}$ pairs separately; we did not require them to be in a singlet. Their triplet combination is a spin rotation of $\ket{\psi_{0,1}} = t^\dagger \ket{\text{vac.}}$ and is also annihilated by $H_{\text{kin}, a}$.

We remark that we numerically observe other entanglement entropy outlier states. Specifically, we observe states with the same energy and in the same symmetry sector as $\ket{\phi_N}$ which appear to contain $N$ doublons and an $X$ dependent superposition of a long-range entangled singlet. There are also entanglement entropy outliers at $k\neq 0,\pi$ and $S=0$. However, these states are not states of well defined doublon number and do not have simple energies. A detailed study of these outlier states could be interesting future work. 

Lastly, we note that the terms \#6,7,8, and \#5,11 (with $\phi=0$) in Table~\ref{tab:operatorstranslate} preserve both scar towers. Additionally, the towers $\ket{\phi_N}$ and $\ket{\psi_{N,1}}$ are preserved by \#2-4, and both SU(2)-invariant choices of \#1. 
Unlike the eta-pairing states, since these new states involve pairs across bonds, hopping terms ``perpendicular" to these pairs do not cancel, and so we do not expect them to generalize to higher dimensions.

\subsection{Analogous new scar tower in the spin-1 XY model}
Given the similarities identified between the spin-1 XY and Hubbard models, it might not be surprising that in the 1D spin-1 XY model, there is a tower of states analogous to the ``singlet" tower $\ket{\phi_N}$. Specifically, defining a state 
\begin{equation}
\ket{\mathcal{S}^{00}} = \sum_j (-1)^j \ket{-1,...,-1,0_j,0,-1,...,-1} ~,    
\end{equation}
the states 
\begin{equation}
    \ket{\mathcal{S}^{00}_N} = (Q^\dagger)^N \ket{\mathcal{S}^{00}}
\end{equation}
are also zero-energy eigenstates of $H_{XY}$. Similar to the ``singlet" and ``triplet" scar towers in the Hirsch model, this can be proven using the commutator argument and commutation relation~\cite{mark_unified_2020}
\begin{equation}
[H_{XY},Q^\dagger] = J\sum_j (-1)^j (\ketbra{0,1}{-1,0} - \ketbra{1,0}{0,-1})_{j,j+1} ~. 
\end{equation}
Since the 0 only occurs in a single `00' in all of these states, we can write
\begin{align}
\ketbra{0,1}{-1,0}_{j,j+1}\ket{\mathcal{S}^{00}_N} &= \ketbra{0,1,0}{-1,0,0}_{j,j+1,j+2}\ket{\mathcal{S}^{00}_N} ~, \nonumber\\
\ketbra{1,0}{0,-1}_{j,j+1}\ket{\mathcal{S}^{00}_N} &= \ketbra{0,1,0}{0,0,-1}_{j-1,j,j+1}\ket{\mathcal{S}^{00}_N} ~.
\end{align}
This gives
\begin{align}
&[H_{XY},Q^\dagger]\ket{\mathcal{S}^{00}_N} = \\
&J\sum_j (-1)^j \ket{0,1,0}(\bra{-1,0,0} + \bra{0,0,-1})_{j,j+1,j+2}\ket{\mathcal{S}^{00}_N}~.\nonumber 
\end{align}
The last line is zero because the `00' in $\ket{\mathcal{S}^{00}_N}$ has momentum $\pi$. The commutator argument is complete by noting that $H_{XY}$ annihilates the base of the tower $\ket{\mathcal{S}^{00}}$, again because the `00' has momentum $\pi$. This tower is similar to the ``singlet" tower in that only one `00' can be present in the tower of states. Accordingly, there is no analogue to the ``triplet" towers of states $\ket{\psi_{N,M}}$ in the spin-1 XY model. We lastly note that due to special symmetries present in the 1D spin-1 XY model~\cite{schecter_weak_2019,chattopadhyay_quantum_2019}, we have to add a longer range term such as $\sum_j (\ket{0,0,1}+\ket{1,0,0})(\bra{0,0,1}+\bra{1,0,0})_{j-1,j,j+1}$ for this tower of states to be ``true scars," that is, states in a quantum chaotic spectrum.

\section{Systematic construction of spin models that commute with $Q^\dagger$}
\label{sec:AppendixSystematicComm}
In this Appendix we obtain the set of nearest-neighbor operators that commute with $Q^\dagger$~(Section~\ref{subsec:opsthatcommute}).
Restricting to Hermitian operators $O$, $[O, Q^\dagger] = 0$ implies that $[O, S^z_\text{Tot.}] = 0$.
We express the operators $O=\sum_j o_j$ in terms of the two-site bases $\ketbra{a,b}{c,d}_{j,j+1}$, $a+b=c+d$. It suffices to consider the commutator $[o_j,\big((S^+_j)^2 - (S^+_{j+1})^2\big)/2] = [o_j, q^\dagger_j]$. $q^\dagger_j$ connects sectors $S^z_j + S^z_{j+1} = -2,0,2$ and $S^z_j + S^z_{j+1} = -1,1$, so we can consider these groups separately.

The $S^z_j + S^z_{j+1} = -2,0,2$ group gives:
\begin{align}
    o_j = &\begin{pmatrix}
    a \\
    & b & c  &d\\
    & c^* & e & f\\
    & d^* & f^* & g\\
    & &&& h
    \end{pmatrix}_{j,j+1} 
    \begin{matrix}
    \ket{1,1}\\
    \ket{1,-1}\\
    \ket{0,0}\\
    \ket{-1,1}\\
    \ket{-1,-1}
    \end{matrix}~, \label{eq:commmatrix}\\
    q^\dagger_j = 
     &\begin{pmatrix}
    0&-1& 0 & 1 & 0 \\
     &&&& 1\\
    &&&& 0\\
    &&&& -1\\
    &&&&0 
    \end{pmatrix}_{j,j+1}~.
\end{align}
The matrix basis is indicated in Eq.~(\ref{eq:commmatrix}) --- for example the ``$d$" entry indicates the term $d \ketbra{1,-1}{-1,1}_{j,j+1}$. Evaluating the commutator, we get:
\begin{align}
    &[o_j, q^\dagger_j] =\label{eq:commutatormat}\\ &\begin{pmatrix}
    0 ~ & b-d^*-a ~ & c - f^* ~ & a+d-g ~ & 0\\
    &&&& b-d-h\\
    &&&& c^*-f\\
    &&&& d^* + h - g\\
    &&&& 0 
    \end{pmatrix}_{j,j+1}. \nonumber
\end{align}
For $[O,Q^\dagger] = 0$, each matrix entry must be zero. We obtain the linearly independent operators \#1,2,6,7 in Table~\ref{tab:operators}.

Performing a similar analysis for the $S^z_j + S^z_{j+1} = -1,1$ group gives the terms \#3,4,5 in Table~\ref{tab:operators}.

\section{Analytic proof of spin-1 XY scarred models}
\label{app:analyticSpin1Scarry}
In Sec.~\ref{sec:spinmodels}, we performed a brute force numerical search to find the family of all translationally invariant nearest-neighbor models that contain the spin-1 XY scar tower (Table~\ref{tab:operators}). While the numerical search can be generalized to other scar towers, such as the AKLT scar towers in Appendix~\ref{sec:AKLTfamily}, in this appendix we analytically prove our result in Sec.~\ref{sec:spinmodels}.

It will be convenient to use the following basis for spin-1, two site states~\cite{mark_unified_2020}:
\begin{align}
& \ket{X_1} = (\ket{1,-1} + \ket{-1,1})/\sqrt{2}~,~~~ \ket{X_2} = \ket{0,0}, \nonumber\\
& \ket{X_3} = \ket{1,0},~~~ \ket{X_4} = \ket{0,1}, \nonumber \\
& \ket{X_5} = \ket{-1,0},~~~ \ket{X_6} = \ket{0,-1},\label{eq:ketXs} \\
& \ket{X_7} = (\ket{1,-1} - \ket{-1,1})/\sqrt{2}~, \nonumber\\
& \ket{X_8} = \ket{1,1},~~~ \ket{X_9} = \ket{-1,-1}. \nonumber
\end{align}
We are interested in finding operators that annihilate all $\ket{\mathcal{S}_N}$. We first note that $\ketbra{...}{X_i}$ for $1 \leq i \leq 6$, trivially annihilates the scar tower $\ket{\mathcal{S}_N}$. So it suffices to consider the action of $\ketbra{...}{X_i}$, $7 \leq i \leq 9$ on $\ket{\mathcal{S}_N}$.

As in Table~\ref{tab:operators}, we restrict our attention to terms that preserve $S^z_\text{Tot.}$. The terms $\sum_j \ketbra{X_8}{X_8}_{j,j+1}$ and $\sum_j \ketbra{X_9}{X_9}_{j,j+1}$ do not annihilate the ferromagnetic states $\ket{\mathcal{S}_L} = \ket{1,1,...,1}$ and $\ket{\Omega} = \ket{-1,-1,...,-1}$ respectively (and are the only such terms, so their action cannot be cancelled out). Therefore we only need to consider terms in the linear space $\sum_j \{\ketbra{X_1}{X_7}, \ketbra{X_2}{X_7}, \ketbra{X_7}{X_7} \}_{j,j+1}$. Just by considering annihilation of the first state of the tower, $\ket{\mathcal{S}_1} = Q^\dagger \ket{\Omega}$, we conclude that the only Hermitian term we can construct from this space and their Hermitian conjugates are: $\sum_j c\ketbra{X_1}{X_7}_{j,j+1} + \hc$. The choices $c=1$ and $c=i$ span this space. The $c=1$ choice is in fact related to simpler terms $\sum_j \left(\ketbra{X_4} - \ketbra{X_3}\right)_{j,j+1}$ (see Eq.~(\ref{eq:sum1j1j+1}) in Sec.~\ref{subsubsec:nullops}), while the $c=i$ choice is proportional to \#12 in Table.~\ref{tab:operators}. That \#12 annihilates all scar states $\ket{\mathcal{S}_N}$ was proven in Sec.~\ref{sec:discususionofresults}.

This exhausts the space of nearest-neighbor, translationally invariant operators that annihilate $\ket{\mathcal{S}_N}$, and we recover the results in Table~\ref{tab:operators}.

\section{Complete family of spin-1 models scarred with $k=0$ bimagnon tower}
\label{sec:q0towers}
The analytical argument in Appendix~\ref{app:analyticSpin1Scarry} can also be used to find all models that are scarred by the ``$k=0$ bimagnon tower." By this we mean the tower of states
\begin{gather}
    \ket{\mathcal{S}_N^{k=0}} = (Q_{k=0}^\dagger)^N \ket{\Omega}~, \\
    Q_{k=0}^\dagger = \frac{1}{2} \sum_j (S^+_j)^2~,~~~ \ket{\Omega} = \ket{-1,...,-1}~.
\end{gather}
The $Q^\dagger_{k=0}$ is similar to the $Q^\dagger$ operator in the main text ($Q^\dagger \equiv Q^\dagger_{k=\pi}$), except that it imparts zero momentum, instead of momentum $\pi$. Given that two distinct models --- the spin-1 XY and spin-1 AKLT models --- contain towers related by $Q^\dagger$, it is natural to ask if there are any physically interesting models that host the $k=0$ bimagnon tower of states. We find that, without one special term, any nearest-neighbor model containing the $k=0$ bimagnon tower conserves the number of 0s $n_0$, and hence can be mapped, in the $n_0 = 0$ symmetry sector, to a family of spin-1/2 models with arbitrary Heisenberg interactions and special Dzyaloshinskii-Moriya-type interaction.

Repeating the argument in Appendix~\ref{app:analyticSpin1Scarry}, we can analytically find all nearest-neighbor models that annihilate $\ket{\mathcal{S}^{k=0}_N}$. Using the notation in Eq.~(\ref{eq:ketXs}), we first note that any term $\ketbra{...}{X_i}$ for $2\leq i \leq 7$ annihilates $\ket{\mathcal{S}^{k=0}_N}$. For $S^z_\text{Tot.}$-preserving Hamiltonians, besides the above trivially-annihilating terms, we only need to consider terms $\sum_j \ketbra{X_8}_{j,j+1},~\sum_j \ketbra{X_9}_{j,j+1}$ and in the linear space $\sum_j \{\ketbra{X_1}{X_1}, \ketbra{X_2}{X_1}, \ketbra{X_7}{X_1} \}_{j,j+1}$. As in the $k=\pi$ bimagnon case, we can prove that the only Hermitian terms we can construct using this set that annihilate the $k=0$ bimagnon tower are: $\sum_j c \ketbra{X_1}{X_7} + \hc$.
Of these, only the $c=i$ choice gives a term independent of the previously considered terms $\ketbra{...}{X_i}$ with $2\leq i \leq 7$. Thus, the term
\begin{align}
& i\sum_j \left(\ketbra{X_1}{X_7} - \ketbra{X_7}{X_1}\right)_{j,j+1} \\
& = i\sum_j \left(\ketbra{-1,1}{1,-1} - \ketbra{1,-1}{-1,1} \right)_{j,j+1} \\
& = \frac{i}{4} \sum_j \left[(S_j^-)^2 (S_{j+1}^+)^2 - (S_j^+)^2 (S_{j+1}^-)^2 \right] \label{eq:term12spin1}
\end{align}
is the only new term that annihilates the $k=0$ bimagnon tower, and unlike the other terms the mechanism of annihilation is non-local.

In this set of operators, there is only one term that does not conserve the number of zeroes: 
\begin{equation}c\ketbra{X_2}{X_7} + \hc = c\ket{0,0}(\bra{1,-1} - \bra{-1,1})/\sqrt{2} + \hc~.\label{eq:q0preserving}
\end{equation}
This term is anti-symmetric under spatial inversion and in general more difficult to come by than the $k=\pi$ case. For the $k=\pi$ bimagnon tower, the corresponding term is $c\ket{0,0}(\bra{1,-1} + \bra{-1,1}) + \hc$, which is present in the $H_{XY}$ spin exchange term, see Eq.~(\ref{eq:XYdecomp}).

However, Eq.~(\ref{eq:q0preserving}), with $c=i$, can arise as part of a natural-looking Dzyaloshinskii–Moriya-type interaction (DMI)
\begin{align}
&h^\text{DMI}_{j,j+1} = \mathbf{\hat{z}} \cdot (\mathbf{S}_j \times \mathbf{S}_{j+1}) = S_j^x S_{j+1}^y - S_j^y S_{j+1}^x \\
&= i(\sqrt{2} \ketbra{X_7}{X_2} + \ketbra{X_3}{X_4} + \ketbra{X_6}{X_5})_{j,j+1} + \hc ~,\nonumber
\end{align}
where the additional parts also annihilate the $k=0$ bimagnon tower.
On the other hand, the case $c=1$ corresponds to a less natural-looking term
\begin{align}
i (S_j^x S_{j+1}^y - S_j^y S_{j+1}^x) S_j^z S_{j+1}^z + \hc \sim \ketbra{X_7}{X_2} + \hc ~, \nonumber
\end{align}
which in addition breaks physical time reversal invariance and will not be considered further. On a 1D chain with only nearest-neighbor interactions, the former DMI model is in fact unitarily related to the XY chain of Schecter and Iadecola~\cite{schecter_weak_2019}, by the transformation $H_{XY} = U H_\text{DMI} U^{-1}$, with
\begin{equation}
    U = \prod_j \exp\left(i\frac{\pi}{2}j S^z_j \right) = \otimes_j \begin{pmatrix} i^{j} &0 &0 \\
    0 & 1 & 0 \\
    0 & 0 & (-i)^j
    \end{pmatrix}_j~,
\end{equation}
in open boundary conditions (OBC) and periodic boundary conditions (PBC) for $L = 4n$. The same unitary relates also the $k=0$ and $k=\pi$ bimagnon towers: $U \ket{\mathcal{S}^{k=0}_N} \propto \ket{\mathcal{S}_N}$. (For PBC in $L = 4n + 2$, $H_\text{DMI}$ rotates to an XY chain with antiperiodic boundary conditions. Because the proof for these terms to annihilate the XY scar tower in Sec.~\ref{sec:spinmodels} relied on strictly local annihilation, the antiperiodic XY model, with flipped sign on $-S^x_L S^x_1 - S^y_L S^y_1$, also contains the $k=\pi$ bimagnon tower.) 

Furthermore, thinking directly about models with $k=0$ bimagnon towers, it is clear that the above term does not require bipartite structure and in general is not simply derived from the spin-1 XY model.

To summarize, analogously to how the model $H_{XY} + h S^z_\text{Tot.} + D \sum_\mathbf{j} (S^z_\mathbf{j})^2$ [Eq.~(\ref{eq:HIS})] hosts the $k=\pi$ tower of scars in any dimension, we expect the model $H_\text{DMI} + h S^z_\text{Tot.} + D \sum_\mathbf{j} (S^z_\mathbf{j})^2$ to host the $k=0$ tower of scars in any dimension. In 1D, just as we need to introduce the additional term $H_3 = J_3 \sum_j (S^x_j S^x_{j+3} + S^y_j S^y_{j+3})$ to break additional symmetries, we need to likewise introduce a range-$k$ DMI term $H'_k = \sum_j (S^x_j S^y_{j+k} - S^y_j S^x_{j+k})$ to break the equivalent symmetry (because we do not need bipartiteness, $k$ can be even or odd here). 

Finally, the term in Eq.~(\ref{eq:term12spin1}) can also be generalized to higher dimensions, by placing it on oriented loops (maintaining the same coupling along the loop), exactly as in the generalization of the term \#12 to higher dimensions in Sec.~\ref{sec:discususionofresults}.

\subsubsection{Application: Generalization of models with $\pi$-bimagnon towers to arbitrary graphs}

In the main text, we considered models realizing $\pi$-bimagnon towers on bipartite lattices, with two-site terms defined only on links connecting different sublattices A and B.  We can immediately generalize these to models with two-site terms defined also on links connecting sites on the same sublattice, A-A or B-B.  Indeed, from the point of view of one such sublattice, the $\pi$-bimagnon states look like the $k=0$ bimagnon states considered in this Appendix.  Hence, all terms considered above, placed on either A-A or B-B links, will preserve the $\pi$-bimagnon tower of Iadecola and Schecter.

\subsubsection{Application: Reduction to new spin-1/2 models with a magnon scar tower} 

In the absence of the $c\ketbra{X_2}{X_7} + \hc$ term [Eq.~(\ref{eq:q0preserving})], a model containing the $k=0$ bimagnon scar tower will preserve the number of 0s $n_0$. The scar tower will lie in the $n_0 = 0$ sector. In this sector, the only relevant terms are those that only involve 1s and -1s. These terms can be mapped onto a spin-1/2 model, replacing 1 with $\up$ and -1 with $\dn$.

There are three terms in the identified family that contain only 1s and -1s. They are: 
\begin{gather}
\ketbra{X_7}_{j,j+1} \mapsto \big(\ketbra{\up,\dn} + \ketbra{\dn,\up} \label{eq:spin12heis}\\
- \ketbra{\up,\dn}{\dn,\up}- \ketbra{\dn,\up}{\up,\dn}\big)_{j,j+1} = \frac{1}{2} \left( I_{j,j+1} - \mathbf{s}_j \cdot \mathbf{s}_{j+1} \right)~, \nonumber\\
\sum_j \ketbra{X_1}{X_7}_{j,j+1} + \hc \mapsto \sum_j (\ketbra{\up\dn} - \ketbra{\dn\up})_{j,j+1}\nonumber\\
 = 0~, \label{eq:actually0}\\
i\sum_j \ketbra{X_1}{X_7}_{j,j+1} + \hc \mapsto i\sum_j (s^-_j s^+_{j+1} - s^+_j s^-_{j+1}) \nonumber\\
 = 2\sum_j (s^y_j s^x_{j+1} - s^x_j s^y_{j+1})~, \label{eq:DMI}
\end{gather}
where we use $\mathbf{s}_j$ to denote spin-1/2 spin operators. Eq.~(\ref{eq:actually0}) can be seen because $\sum_j \ketbra{X_1}{X_7}_{j,j+1} + \hc$ maps to a counting of opposite domain walls $N_{\up\dn} - N_{\dn\up} = 0$, which is zero in a closed spin-1/2 chain. The term in Eq.~(\ref{eq:spin12heis}) corresponds to a spin-1/2 Heisenberg model, while the term in Eq.~(\ref{eq:DMI}) corresponds to a Dzyaloshinskii-Moriya interaction (DMI) of spin-1/2's and also corresponds to the term \#12 in Table~\ref{tab:operators}. 

Therefore, if the $c\ketbra{X_2}{X_7} + \hc$ term is not present, a spin-1 model scarred by the $k=0$ bimagnon tower is equivalent to, in the $n_0 = 0$ sector, a spin-1/2 Heisenberg model with Dzyaloshinkii-Moriya interaction. On the 1D chain with only nearest-neighbor terms, this model is integrable~\cite{alcaraz_heisenbergxxz_1990}, by essentially undoing the spin ``twist" in the DMI and transforming to an XXZ chain with twisted boundary conditions, which is in turn solvable by the Bethe ansatz.

The spin-1 model $k=0$ bimagnon tower maps onto the simple spin-1/2 model $k=0$ magnon tower, generated by repeated action of the operator $\sum_j s_j^+$ on the state $\ket{\dn,\dn,\dots,\dn}$.  This tower is nothing else but the familiar $S=\text{Vol}/2$ highest-spin multiplet which would describe the degenerate ferromagnetic states in an SU(2)-invariant model.

We conclude from these observations that any inversion symmetric, nearest neighbor spin-1 model cannot contain the spin-1 $k=0$ bimagnon tower as scars: it will possess $n_0$ conservation and be equivalent to the spin-1/2 Heisenberg model in the $n_0=0$ sector. The bimagnon states are not scars because of the spin-1/2 SU(2) symmetry in the $n_0=0$ sectors.

However, there are interesting spin-1/2 models scarred by the $k=0$ magnon tower. These states will be eigenstates of any spin-SU(2) invariant model, for example the $J_1$-$J_2$ model 
\begin{equation}
    H_{J_1-J_2} = \sum_j ( J_1 \mathbf{s}_j \cdot \mathbf{s}_{j+1} + J_2 \mathbf{s}_j \cdot \mathbf{s}_{j+2} )~.
\end{equation}
We can then add the spin-1/2 DM interaction to get a simple non-integrable spin-1/2 model scarred by the $k=0$ magnon tower: 
\begin{equation}
H_{J1-J2} + D H_\text{DMI}~,~~~H_\text{DMI} = \sum_j \mathbf{\hat{z}} \cdot (\mathbf{s}_j \times \mathbf{s}_{j+1}) ~.
\end{equation}
This model has been considered in the context of the magneto-electric effect in ferroelectric materials~\cite{katsura_spin_2005,vahedi_1d_2012,brockmann_exact_2013}. The DM interaction breaks the spin SU(2) symmetry, and the presence of this multiplet is a non-trivial ``scar" property.

We verified numerically that this model in 1D is indeed scarred by the $k=0$ magnon scar tower. Apart from translational invariance and $S^z$ conservation, there is an additional spin flip plus inversion symmetry given by $g=\prod_j \sigma^x_j \times I$, where $I\ket{s_1, s_2, \dots, s_L} = \ket{s_L, s_{L-1}, \dots, s_1}$. However, for small $D$, the couplings $J_1$ and $J_2$ must have opposite signs in order for the ferromagnetic states to be in the bulk of the spectrum. In our numerical study with couplings $J_1=1, J_2=-0.6, D=0.3$, this was indeed the case for the scar state in the sector $k=0, S^z_\text{Tot.}=0, g=1$. On the other hand, $J_1$ and $J_2$ are usually assumed to have the same sign for the purposes of frustrated antiferromagnetism, in which case the eigenstates considered here would be all ceiling states and nominally not scar states, although the fact that they are all degenerate eigenstates is still non-trivial.

We note that the same scar states were also obtained in a Shiraishi-Mori type ``toy model" in Ref.~\cite{Choi2019} with complicated four-spin interactions.  It is remarkable that we found a much simpler and more realistic model with only two-spin interactions; also, our model does not appear to be of Shiraishi-Mori type with two-site projectors.

This model can be extended to higher dimensions, as long as the DMI terms occur in loops. That is, orienting the DMI vectors with $+\hat{z}$, each directed bond $\mathbf{i} \rightarrow \mathbf{j}$ in the term $s^y_\mathbf{i} s^x_\mathbf{j} - s^x_\mathbf{i} s^y_\mathbf{j}$ belongs to a unique closed loop $\mathbf{i} \rightarrow \mathbf{j} \rightarrow \mathbf{k} \cdots \rightarrow \mathbf{i}$. This is in fact the case for the DMI studied in the kagome lattice material Herbertsmithite~\cite{Zorko_DM_2008, Rousochatzakis_DM_2009,lee_gapless_2018} and the triangular lattice material Cs$_2$CuCl$_4$ studied in Ref.~\cite{coldea_direct_2002}, for example. Again, if all nearest-neighbor and further-neighbor Heisenberg couplings are antiferromagnetic, we expect that for small DMI these eigenstates are ceiling states and nominally not scars, but they are still special.

We also note that we can easily prepare a suitable initial state for perfect revivals in such scarred systems. Following Iadecola and Schecter's initial state for the spin-1 XY scars, we can give the ferromagnetic scars an equal energy splitting with the term $h S^z_\text{Tot.}$. The ferromagnetic state in the $x$ direction $\otimes_j \left[(\ket{\up} + \ket{\dn})_j /\sqrt{2} \right]$ is a superposition of the $k=0$ magnon states and hence will experience perfect revivals.

Finally, we can also consider spin-1/2 models that contain the spin-1/2 $k=\pi$ magnon tower. The DMI term corresponds to \#12 in Table~\ref{tab:operators} and also annihilates the $k=\pi$ magnon tower. Instead of the Heisenberg term, we we need to consider $\sum_j \ketbra{X_1}_{j,j+1}$ instead, which corresponds to a less-natural XXZ model $\sum_j (s^x_j s^x_{j+1} + s^y_j s^y_{j+1} - s^z_j s^z_{j+1})$. Therefore, in the spin-1/2 context, it is more natural to consider models scarred by the $k=0$ magnon tower, instead of the $k=\pi$ one.

\section{Exhaustive search for models scarred by a given scar tower}\label{sec:exhaustivesearch}

In this appendix we provide a framework to exhaustively find all models that contain a given scar tower, provided the scar tower satisfies certain conditions. We then apply this method in Appendix~\ref{sec:AKLTfamily} to find a complete family of nearest-neighbor models scarred with the AKLT scars.

We are given a scar tower $\{\ket{\mathcal{S}_N}\}$, obtained by $\ket{\mathcal{S}_N} = \left(Q^\dagger\right)^N \ket{\mathcal{S}_0}$, for some operator $Q^\dagger$ and state $\ket{\mathcal{S}_0}$. We then want to find a family of all models such that the $\ket{\mathcal{S}_N}$ are eigenstates with energy $E_N = qN + \lambda$, for some $q$ and $\lambda$. Without loss of generality, we can set $\lambda=0$ by subtracting $\lambda I$. We observe that for any such model $H$, we have that:
\begin{equation}
    \forall N~,~ \left[H, Q^\dagger \right] \ket{\mathcal{S}_N} = q Q^\dagger \ket{\mathcal{S}_N}~,~ H \ket{\mathcal{S}_0} = 0~.
\end{equation}
Suppose we have $H_Q$ such that
\begin{equation}
    \left[H_Q, Q^\dagger \right] = Q^\dagger~,~ H_Q \ket{\mathcal{S}_0} = 0 ~.
\end{equation}
We can then split $H$ into parts:
\begin{equation}
    H = q H_Q + H'~.
\end{equation}
where: 
\begin{equation}
    \forall N~,~ \left[H', Q^\dagger \right] \ket{\mathcal{S}_N} = 0 ~
\label{eq:hamdecomp}
\end{equation}
In our case of interest, $Q^z = S^z_\text{Tot.}/2$ satisfies $\left[Q^z, Q^\dagger\right] = Q^\dagger$. Furthermore, in our examples, $\ket{\mathcal{S}_N}$ are eigenstates of $Q^z$.

It follows that $\ket{\mathcal{S}_0}$ is also an eigenstate of $H'$ (whose eigenvalue we can again set to 0). Then Eq.~(\ref{eq:hamdecomp}) is equivalent to:
\begin{equation}
   \forall N~,~ H' \ket{\mathcal{S}_N} = 0~.
\end{equation}

It suffices to consider operators $H'$ that change the magnetization $S^z_\text{Tot.}$ by a fixed amount. This is because we require $H'$ to annihilate each $\ket{S_N}$ individually. Since each $\ket{S_N}$ has fixed magnetization, if $H'$ had components changing $S^z_\text{Tot.}$ by a different amount, the images, upon action of different such components, would be in different $S^z_\text{Tot.}$ sectors and would have to independently cancel, i.e., each component independently annihilates all $\ket{S_N}$.

Restricting to operators that change $S^z_\text{Tot.}$ by a fixed amount has the following advantage: searching for all such $H'$ is equivalent to finding all Hamiltonians that annihilate a compression of the tower states $\sum_N c_N \ket{\mathcal{S}_N}$. In the case of the AKLT model scar tower discussed below, this compression can be written as an MPS, allowing us to use the methods discussed in Sec.~\ref{subsec:opsthatannihilate}. We lastly note that the identity should be trivially added to this list $H'$, since it was used to set all eigenvalues to 0.

\section{Complete family of nearest-neighbor models scarred with AKLT scars}
\label{sec:AKLTfamily}
Our systematic search for models scarred with the spin-1 XY scar tower revealed the new term \#12 that cannot be reduced to local terms annihilating these scar states.  Motivated by this observation, in this appendix we derive the complete family of nearest-neighbor models that contain the AKLT tower of scar states.  These scar states are defined by acting the same operator $Q^\dagger$ in Eq.~(\ref{eq:ISXYscar1}) on the AKLT ground state~$\ket{G}$.  It can be compactly expressed by the MPS~\cite{moudgalya_entanglement_2018}:
\begin{equation}
    \ket{G} = \sum_{\{\sigma_1 \cdots \sigma_L\}} \Tr(A^{[\sigma_1]} \cdots A^{[\sigma_L]}) \ket{\sigma_1 \cdots \sigma_L} ~,
\end{equation}
where
\begin{align}
    A^{[-1]} &= \sqrt{\frac{2}{3}} \begin{pmatrix} 0 & 1\\ 0 & 0 \end{pmatrix} ~, 
    \quad A^{[0]} = \frac{1}{\sqrt{3}} \begin{pmatrix} -1 & 0\\ 0 & 1 \end{pmatrix} ~, \nonumber \\ 
    A^{[1]} &= \sqrt{\frac{2}{3}} \begin{pmatrix} 0 & 0\\ -1 & 0 \end{pmatrix} ~.
\label{eq:MPS}
\end{align}
The scar states in the AKLT model have energies $E_n = 2n$ and are given by~\cite{Moudgalya2018}:
\begin{equation}
    \ket{\mathcal{S}_{n}} = (Q^\dagger)^n \ket{G}~.
\end{equation}

For the AKLT model, it is convenient to work in the basis of states $\{\ket{T_{J,M}}\}$ of well-defined total spin $J$ and magnetization $M$ across two sites~\cite{mark_unified_2020}:
\begin{gather}
    \ket{T_{2,-2}} = \ket{-1,-1}~,~~ \ket{T_{2,-1}} = \frac{1}{\sqrt{2}} \left(\ket{0,-1} + \ket{-1,0} \right)~, \nonumber\\
    \ket{T_{2,0}} = \frac{1}{\sqrt{6}}\left(\ket{1,-1} + 2\ket{0,0} + \ket{-1,1} \right)~, \nonumber\\
    \ket{T_{2,1}} = \frac{1}{\sqrt{2}}\left(\ket{1,0} + \ket{0,1} \right)~,~~ \ket{T_{2,2}} = \ket{1,1}, \nonumber\\
    \ket{T_{1,-1}} = \frac{1}{\sqrt{2}}\left(\ket{0,-1} - \ket{-1,0}\right)~, \label{eq:twositestates}\\
    \ket{T_{1,0}} = \frac{1}{\sqrt{2}}\left(\ket{1,-1} - \ket{-1,1} \right)~, \nonumber\\
    \ket{T_{1,1}} = \frac{1}{\sqrt{2}}\left(\ket{1,0} - \ket{0,1} \right)~, \nonumber\\
    \ket{T_{0,0}} = \frac{1}{\sqrt{3}}\left(\ket{1,-1} - \ket{0,0} + \ket{-1,1} \right) ~.\nonumber
\end{gather}

As discussed in Appendix~\ref{sec:exhaustivesearch}, it suffices to find all terms that annihilate $\ket{\mathcal{S}_n}$. To do so we compress all $\ket{\mathcal{S}_n}$ into a single MPS with the MPO in Eq.~(\ref{eq:MPO}), then use the MPS method outlined in Sec.~\ref{subsec:opsthatannihilate}. 
By considering translationally invariant nearest-nighbor operators, as in the main text, we found that the following terms annihilate $\ket{\mathcal{S}_n}$:
\begin{gather}
    H_0 = \sum_j \left[\left(\ketbra{0,1}{1,0} - \ketbra{0,-1}{-1,0} \right)_{j,j+1} +\hc \right] ~, \nonumber\\
    \ketbra{T_{2,-2}}_{j,j+1},~\ketbra{T_{2,-1}}_{j,j+1},~\ketbra{T_{2,0}}_{j,j+1},\nonumber\\
    e^{i\phi}\ketbra{T_{2,-2}}{T_{2,-1}}_{j,j+1}+\hc~,~e^{i\phi}\ketbra{T_{2,-2}}{T_{2,0}}_{j,j+1}+\hc~, \nonumber\\
    e^{i\phi}\ketbra{T_{2,-1}}{T_{2,0}}_{j,j+1}+\hc~. \label{eq:d6}
\end{gather}
As in the spin-1 XY model case, most of these terms annihilate $\ket{\mathcal{S}_n}$ bondwise, as already written above. Unlike the spin-1 XY model case, however, the $H_0$ term needs its sum over $j$. In addition, $H_0$ and $\ketbra{T_{2,0}}_{j,j+1}$ commute with $Q^\dagger$. However, since the scar states here do not have a special relationship with the pseudospin symmetry (namely, they are not eigenstates of the total pseudospin),
this property of the two terms is less important~\cite{mark_unified_2020}. To get the scar energies $E_n = 2n$ we can add $S^z_\text{Tot.}$. In fact, the AKLT model is contained in this space, by the fact that~\cite{mark_unified_2020}:
\begin{align}
    \frac{1}{2}H_0+S^z_\text{Tot.} = &\sum_j \big(\ketbra{T_{2,2}} + \ketbra{T_{2,1}} \\
    &-\ketbra{T_{2,-1}} - \ketbra{T_{2,-2}}\big)_{j,j+1} \nonumber \\
    = &H_{AKLT} -  \sum_j \big(\ketbra{T_{2,0}} \nonumber\\
    &+2\ketbra{T_{2,-1}} + 2\ketbra{T_{2,-2}} \big)_{j,j+1}~. \nonumber
\end{align}
We can then express the family of Hamiltonians scarred by the AKLT tower as:
\begin{equation}
H_{AKLT} + h S^z_\text{Tot.} + \sum_j \!\sum_{m,n =-2}^0 \!\!\!\! c_{m,n}(j) \ketbra{T_{2,m}}{T_{2,n}}_{j,j+1} ~,
\end{equation}
reproducing the results from Refs.~\cite{mark_unified_2020,moudgalya_large_2020}.

\end{document}